\documentclass[preprint,12pt]{elsarticle}

\usepackage{amssymb}

\journal{Physics of the Dark Universe}

\begin{document}

\begin{frontmatter}



\title{
Exploring the Role of 
Axions and Other WISPs\\ in the Dark Universe}


\author{Andreas Ringwald}

\address{Deutsches Elektronen-Synchrotron DESY, Notkestr. 85, D-22607 Hamburg, Germany}

\begin{abstract}
Axions and other very weakly interacting slim particles (WISPs) may be non-thermally produced 
in the early universe and survive as constituents of the dark universe. We describe their 
theoretical motivation and their phenomenology.   
A huge region in parameter space spanned by their couplings to photons and their masses can 
give rise to the observed cold dark matter abundance. A wide range of experiments --  
direct dark matter searches exploiting microwave cavities, searches for solar axions or
WISPs, and light-shining-through-a-wall searches --  can probe large parts of this parameter space 
in the foreseeable future. 
\end{abstract}




\end{frontmatter}



\section{Introduction}
\label{introduction}

The standard model of particle physics describes the basic properties of matter and forces  
to the extent we have been able to probe thus far. Yet, it is not a complete and 
fundamental theory of nature since it leaves some big questions unanswered. Notably, it does not 
provide satisfactory
explanations for the values of its many parameters, it does not detail a mechanism that reconciles gravity 
with quantum mechanics, and it does not explain the origin of dark energy and dark matter.

The three most relevant features of particle candidates of dark matter indirectly deduced from observations are 
their feeble interactions with standard model particles and with themselves, 
their sufficiently non-relativistic momentum distribution during structure formation, and 
their stability on cosmological time-scales. 

Weakly interacting massive particles (WIMPs) provide a simple model realising all these features. Firstly, their interactions are of electroweak strength due to the large mass of the mediator particles (such as W or Z bosons). Secondly, being 
thermally produced in the early universe, their large mass, of order the electroweak scale,  
ensures that they are sufficiently cold at the epoch when structure formation begins. 
Thirdly, their stability is ensured by invoking symmetries that conserve their particle number. 
However, motivating such symmetries on theoretical grounds is non-trivial: global symmetries may be broken in quantum gravity, whereas local symmetries lead to additional interactions which may cause conflicts with the required weakness of the 
dark matter interactions. Nevertheless, well-motivated WIMP candidates exist, most notably in supersymmetric
extensions of the standard model with R-parity conservation. In fact, in the case that the lightest supersymmetric particle (LSP) is a neutralino -- a mixture of the supersymmetric partners of neutral standard model gauge bosons and Higgses -- 
the latter qualifies for a WIMP.  

The prospects of exploring the electroweak scale with the Large Hadron
Collider (LHC) has understandably focused the dark matter searches on the WIMP paradigm in the last years 
and the experimental community has devoted relatively little effort to explore other possibilities.
Although it is way too early to make a final judgement it is nevertheless noteworthy that the initial searches at the LHC 
as well as direct searches for WIMP dark matter have not given any clear indication of their existence. Because of this 
shortcoming and the mentioned theoretical issues it is interesting and timely to consider alternative ways to realise the essential features of 
dark matter. In this review, we will concentrate on very weakly interacting slim (in the sense of very light) 
particles (WISPs) as well motivated dark matter alternatives to WIMPs.  

In fact, sufficient stability of the dark matter particles can also be achieved by combining the weakness of their interactions with a sufficiently small mass. The latter drastically reduces the phase space and the number and types of possible decay products, thereby increasing the lifetime. Furthermore, there are non-thermal means for producing sufficiently cold dark matter made of light particles, one of the most generic being the vacuum-realignment mechanism, discussed originally in the case of the QCD axion -- a theoretically very well motivated basic paradigm of a WISP because it is predicted in a 
generic framework proposed to solve the so-called strong CP problem, providing a satisfactory explanation why 
the theta-parameter in QCD vanishes within experimental accuracy. Other WISPs, which go under the name 
``axion-like particles" and ``hidden sector photons", are predicted in certain string theory based extensions of the standard 
model, in addition to a QCD axion candidate. They can be produced also via the vacuum-realignment mechanism. 
The feeble strength of the interactions of axions and other WISPs 
typically results from underlying new dynamics at energy scales much larger than the electroweak scale.  

This review is organised as follows. 
The theoretical motivation for the QCD axion and other WISPs and their phenomenological constraints 
arising mainly from astrophysics and cosmology of the early universe are sketched in Sec.~\ref{sec:science}, while Sec.~\ref{sec:cosmology} 
deals with their relevance for the cold dark matter in the universe. An overview of the experimental landscape 
of axion and other WISP searches  
-- an emerging low-energy, high-intensity frontier~\cite{Jaeckel:2010ni,Hewett:2012ns,Baker:2012esg} -- is given in Sec.~\ref{sec:searches}. Finally, in Sec.~\ref{sec:conclusions} 
we summarise and conclude.

\section{Physics case for axions and other WISPs}
\label{sec:science}

In this section we review the physics case for axions and other WISPs. The solution of 
the strong CP problem -- providing an explanation of the non-observation of strong CP violation -- 
gives a particularly strong motivation for the existence of the QCD axion, cf. Sec.~\ref{sec:qcdaxion}. 
There are also good reasons to expect axion-like particles (ALPs) and light hidden sector $U(1)$ gauge 
bosons as well as minicharged particles, 
particularly in extensions of the standard model inspired by string theory, cf. Sec.~\ref{sec:alpsfromstrings}. 
There are hotspots in the landscape of axion, ALP and hidden photon parameters, i.e. 
their masses and their couplings to standard model particles, 
arising from well motivated ultraviolet completions of the standard model. Often, these parameters
can be linked to fundamental scales of particle physics, such as the GUT scale, $M_{\rm GUT}\sim 10^{16}$~GeV, or the 
intermediate scale, $M_{\rm I}\sim \sqrt{M_W M_P}\sim 10^{10}$~GeV, the geometric mean between the weak scale 
$M_W\sim 100$~GeV and the Planck scale $M_P\sim 10^{18}$~GeV, cf. Sec.~\ref{sec:hotspotstheory}. 
From a phenomenological point of view, on the one hand, 
the current limits on the existence of axions and other WISPs from astrophysics and 
early cosmology are already quite stringent, cf. Sec.~\ref{sec:astrophysics}. 
On the other hand, just behind the boundary of current knowledge, 
there are intriguing hints suggesting the existence of axions and 
axion-like particles, compatible with an intermediate scale decay constant, $f_a\sim M_{\rm I}\sim 10^{10}$~GeV, 
cf. Sec.~\ref{sec:astrophysics}.

\subsection{The strong CP problem and the QCD axion}
\label{sec:qcdaxion}

The QCD axion arises in the course of a dynamical solution of the so-called strong 
CP-problem. 
In fact, QCD allows for a CP-violating term in the Lagrangian,
\begin{eqnarray}
{\mathcal L}_{\rm CP-viol.} =
\frac{\alpha_s}{8\pi}\, \theta\, G^a_{\mu\nu} {\tilde G}^{a,\mu\nu} \equiv
\frac{\alpha_s}{8\pi}\, \theta\,
\frac{1}{2}\,\epsilon^{\mu\nu\alpha\beta}\, G^a_{\mu\nu} G^a_{\alpha\beta},
\label{topterm}
\end{eqnarray}
where $G$ is the gluonic field strength. Similar to the strong coupling constant $\alpha_s$,
the fundamental parameter $\theta$ has to be determined experimentally.
One of the most sensitive probes for it is the electric dipole moment $d_n$ of the neutron. It
depends linearly on
\begin{equation}
\bar{\theta} \equiv \theta + {\rm arg\ det\ }M,
\end{equation}
with $M$ being the quark mass matrix. It is of order 
\begin{eqnarray}
\left| d_n\right| \sim \frac{e}{m_n} \left( \frac{m_q}{m_n}\right)
\left|\bar{\theta}\right|
\sim 10^{-16}\ \left|\bar{\theta}\right| \ e\,{\rm cm},
\end{eqnarray}
where $m_n$ ($m_q$) is the neutron (a light-quark mass), $e$ is the unit electric charge.
The current experimental
upper bound on $\left|d_n\right|<2.9\times 10^{-26}\ e\,$cm~\cite{Beringer:1900zz} places an extremely stringent limit,
\begin{equation}
\left|\bar\theta\right| \lesssim 10^{-10}.
\end{equation}
The strong CP problem is the lack of an explanation why the dimensionless parameter $\bar\theta$
is so unnaturally small.

A very well motivated solution was proposed by Peccei and Quinn~\cite{Peccei:1977hh}: their basic idea was to introduce 
the axion field $a$ as a dynamical $\bar\theta$ parameter which can relax spontaneously to zero.  
Assuming that $a$ enjoys a shift symmetry,
\begin{eqnarray}
a\to a + {\rm const.},
\end{eqnarray}
which is broken only by anomalous coupling terms to gauge bosons, in particular to gluons,  
its most general low-energy effective Lagrangian below the weak scale, i.e. where the intermediate vector bosons 
$W$,$Z$, and the Higgs boson are integrated out, can be parametrised 
as~\cite{Georgi:1986df}
\begin{eqnarray}
\mathcal{L} &\supset \frac{1}{2}\, \partial_\mu a\, \partial^\mu a
- \frac{\alpha_s}{8\pi} \left(\bar{\theta}  +  \frac{a}{f_{a}} 
\right) G_{\mu\nu}^b \tilde{G}^{b,\mu\nu}  - \frac{\alpha}{8\pi} C_{a\gamma} \frac{a}{f_{a}} \, F_{\mu\nu} \tilde{F}^{\mu\nu} 
\nonumber\\
&  + \sum_\Psi \bigg[  \overline{\Psi} \gamma^\mu \frac{1}{2}(\tilde{X}_{\psi_R} + \tilde{X}_{\psi_L})\gamma_5   \Psi  +  \overline{\Psi} \gamma^\mu \frac{1}{2}(\tilde{X}_{\psi_R} - \tilde{X}_{\psi_L}) \Psi \bigg] \frac{\partial_\mu a}{f_{a}}
\, .
\label{axion_leff}
\end{eqnarray}
Here, $F$ denotes the electromagnetic field strength, $C_{a\gamma}$ a model dependent dimensionless parameter 
from the electromagnetic anomaly, and $\Psi$ standard model matter fields. 
The dimensionful axion decay constant $f_a$, together with the dimensionless couplings ($C,\tilde X$),  
which are expected to be of order one, 
determine the strength of the interaction of the axion with 
standard model particles. 

The dynamics described by the Lagrangian~(\ref{axion_leff}) solves the strong CP problem 
in the following way. Firstly,  
the $\bar\theta$-term can be eliminated by absorbing it into the axion field, $a=\bar{a} - \bar{\theta}
f_a$. Secondly, the topological charge density $\propto \langle
{\rm tr}\, G^{\mu\nu} {\tilde G}_{\mu\nu} \rangle \neq 0$, induced
by topological fluctuations of the gluon fields such as QCD
instantons, provides a nontrivial potential\footnote{Semiclassical instanton methods 
are not reliable to calculate the potential 
accurately. One has to use matching to the low-energy chiral Lagrangian instead~\cite{Weinberg:1977ma,Georgi:1986df}.} 
for the axion field $\bar{a}$
which is minimized\footnote{The CP invariance of the theory at $\bar{\theta}=0$ guarantees that this is a stationary
point.} at zero expectation value, $\langle \bar{a}\rangle =0$, wiping out strong CP violation. 

The nontrivial potential around $\langle \bar{a}\rangle =0$ promotes the elementary particle excitation of the
axion field, the axion, to a pseudo Nambu-Goldstone
boson~\cite{Weinberg:1977ma,Wilczek:1977pj} (which we will now again denote by $a$) with a non-vanishing, but
parametrically small mass. This mass can be calculated by matching the couplings of the axion to quarks
and gluons onto the appropriate couplings to mesons and baryons in the low-energy effective chiral Lagrangian 
and expressed in terms of the light ($u,d$) quark masses,
the pion mass $m_\pi$ and the pion decay constant $f_\pi$~\cite{Weinberg:1977ma},
      \begin{eqnarray}
m_a =
         \frac{m_\pi f_\pi}{f_a}\frac{\sqrt{m_u m_d}}{m_u+m_d}\simeq { 0.6\,  {\rm meV}}
         \times
         \left(
         \frac{10^{10}\, {\rm GeV}}{f_a}\right) .
         \label{axionmass}
\end{eqnarray}
For large axion decay constant $f_a$, we see that the axion is a particularly well motivated example  
of a WISP: it is indeed slim, i.e. very light, cf. Eq.~(\ref{axionmass}), and it is very weakly 
coupled~\cite{Kim:1979if,Dine:1981rt,Shifman:1979if,Zhitnitsky:1980tq}. In particular, its coupling to photons, which usually is parametrised
as 
\begin{eqnarray}
{\cal L}_{a \gamma \gamma} = - \frac{1}{4}\, g_{a\gamma}\, a\, F_{\mu \nu}\tilde{F}^{\mu \nu} =
g_{a\gamma}\, a\, \vec{E}\cdot \vec{B} ,
\end{eqnarray}
where $\vec{E}$ and $\vec{B}$ denote the electric and magnetic field, respectively, is very small~\cite{Bardeen:1977bd,Kaplan:1985dv,Srednicki:1985xd},
\begin{eqnarray}
        { g_{a\gamma}} = \frac{\alpha}{2\pi f_a}
\left( C_{a\gamma} - {\frac{2}{3}\,\frac{m_u+4 m_d}{m_u+m_d}     }\right)
\sim 10^{-13}\ {\rm GeV}^{-1}          \left(
         \frac{10^{10}\, {\rm GeV}}{f_a}\right).
         \label{axionphotoncoupling}
\end{eqnarray}

Nevertheless, the guaranteed coupling of axions to
photons, Eq.~(\ref{axionphotoncoupling}), may result, if axions
exist, in observable consequences from processes involving large
electromagnetic fields. 
These often occur in astrophysical environments (Sec.~\ref{sec:astrophysics}) 
and can be prepared in laboratory experiments (Sec.~\ref{sec:searches}). 

\subsection{Axion-like particles and hidden sector photons}
\label{sec:alpsfromstrings}

In four dimensional (4D) field theoretic extensions of the standard model, the axion field is 
naturally realised as the phase of a complex $SU(2)_L\times U(1)_Y$ singlet scalar field 
whose vacuum expectation value $v_{\rm PQ}$ breaks a global anomalous chiral $U(1)_{\rm PQ}$ symmetry.
At energies much below 
the symmetry breaking scale $v_{\rm PQ}$, the low-energy effective theory is that of a (pseudo-) Nambu-Goldstone
boson with decay constant $f_a\propto v_{\rm PQ}$, cf. Eq.~(\ref{axion_leff}). This concept has been generalized to other 
-- axion-like particles (ALPs) -- which may arise as (pseudo) Nambu-Goldstone bosons
from the breaking of other global symmetries such as, for example, family symmetries. 

In supersymmetric models which invoke the Peccei-Quinn solution to the 
strong CP problem, an axion supermultiplet must be introduced which
necessarily includes an R-parity even spin-$0$ saxion field $s(x)$ 
and an R-parity odd spin-$1\over 2$ axino field $\tilde{a}(x)$.
They also have couplings suppressed by the decay constant $f_a$.
Nevertheless, they do not qualify as WISPs: they are expected to 
receive TeV scale masses\footnote{ 
The saxion is expected to receive a soft SUSY breaking mass $m_s\sim 1$ TeV.
The axino mass is more model dependent but in gravity-mediated SUSY breaking
models is also expected to gain a mass $\sim 1$ TeV.}.  

A particular strong motivation for the existence of the axion and ALPs comes from string 
theory -- aiming at the currently most ambitious ultraviolet completion of the standard model,
unifying all fundamental forces, including gravity. In fact,  
it has long been known that the 4D low-energy effective field theory emerging string theory  
predicts natural candidates for the QCD axion~\cite{Witten:1984dg,Conlon:2006tq,Svrcek:2006yi},
often even an `axiverse'~\cite{Arvanitaki:2009fg,Acharya:2010zx,Cicoli:2012sz,Ringwald:2012cu} containing
many additional light ALPs whose masses are evenly distributed in log scale. 
Indeed, when compactifying the six extra spatial dimensions of string theory, axions and ALPs arise generically as
Kaluza-Klein zero modes of antisymmetric form fields, the latter belonging to the massless spectrum of the 
bosonic string. Their number is related to the topology of the internal manifold, namely to the number of 
closed submanifolds (``cycles"). 
Given that this number is generically of the order of hundreds,
one expects the low-energy theory to be populated by many axion-like particles\footnote{
These are all closed string axions which live in the bulk. On top of them there can also be open string
axions living on a brane.}. 
Moreover, the anomalous axionic couplings to gauge bosons in their low-energy effective Lagrangian, the 
generalisation of (\ref{axion_leff}) to the case of many ALPs,  
\begin{eqnarray}
\mathcal{L} &\supset \frac{1}{2}\, \partial_\mu a_i\, \partial^\mu a_i
- \frac{\alpha_s}{8\pi} \left(\bar{\theta} + C_{ig} \frac{a_i}{f_{a_i}} 
\right) G_{\mu\nu}^b \tilde{G}^{b,\mu\nu}  - \frac{\alpha}{8\pi} C_{i\gamma} \frac{a_i}{f_{a_i}} \, F_{\mu\nu} \tilde{F}^{\mu\nu} 
\nonumber\\
& 
+ \sum_\Psi \bigg[ \overline{\Psi} \gamma^\mu\frac{1}{2} (\tilde{X}_{\psi_R}^i + \tilde{X}_{\psi_L}^i)\gamma_5   \Psi  +  \overline{\Psi} \gamma^\mu \frac{1}{2} (\tilde{X}_{\psi_R}^i - \tilde{X}_{\psi_L}^i)\Psi \bigg] \frac{\partial_\mu a_i}{f_{a_i}}\, 
,
\label{ALP_leff}
\end{eqnarray}
are also predicted from dimensional reduction of the effective action from ten to four dimensions.
Thus, string compactifications hold the promise of plenty of candidates for the QCD axion and ALPs. 

\begin{figure}[t]
\centerline{\includegraphics[width=0.9\textwidth]{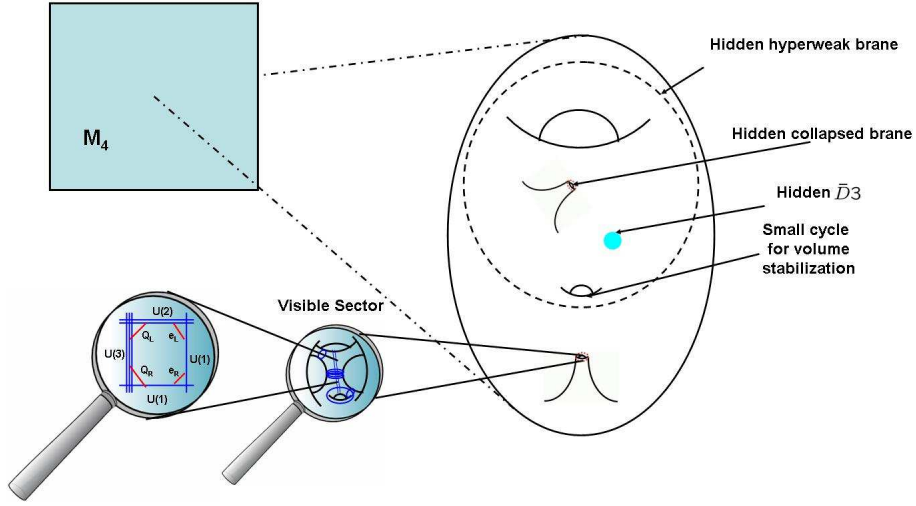}}
\caption{In compactifications of type II string theories the
standard model is locally realised by a stack of space-time filling $D$-branes wrapping
cycles in the compact dimensions (figure from Ref.~\cite{Jaeckel:2010ni}). In
general, there can also be hidden sectors localised in the bulk. 
They can arise from branes of different dimension ($D3$ or $D7$
branes) which can be either of large extent or localised at
singularities. Light visible and hidden matter particles arise from
strings located at intersection loci and stretching between brane
stacks.}
\label{Fig:type_ii_comp}
\end{figure}

Additional hidden sector photons  -- gauge bosons of a  
$U(1)$ gauge group under which the standard model particles are not charged --  also  
generically arise in string compactifications~\cite{Goodsell:2010ie}, providing us with even more well motivated
WISP candidates. In fact, hidden $U(1)$ gauge factors are ubiquitous in string compactifications: In the
heterotic string, they can be found, e.g., in the standard embedding
by breaking the hidden $E_8$ gauge group. In type II string theories, there are so-called RR $U(1)$s, arising
as zero modes of closed string Ramond-Ramond form fields. 
Finally, realising $U(1)$ gauge bosons as massless 
excitations of space-time filling $D$-branes wrapping cycles in the extra
dimensions, extra $U(1)$s can be hidden from the visible sector $U(1)$, if the corresponding
$D$-branes are separated in the compact space (cf. Fig.~\ref{Fig:type_ii_comp}).
In fact, type II compactifications generically 
involve hidden sector branes not intersecting with the standard model branes,
often also for global consistency requirements.

Some of these hidden $U(1)$s may remain unbroken down to very small energy scales.
In this case their dominant interaction with the standard model will be through kinetic mixing with the
photon~\cite{Holdom:1985ag}. The latter is encoded in the low-energy effective Lagrangian,
\begin{equation}
\mathcal{L} \supset -\frac{1}{4} F_{\mu \nu} F^{\mu \nu}
- \frac{1}{4} X_{\mu \nu} X^{\mu \nu}
+ \frac{\chi}{2} F_{\mu \nu} X^{\mu \nu}
+ \frac{m_{\gamma^\prime}^2}{2} X_{\mu} X^{\mu},
\label{LagKM}
\end{equation}
where $X_\mu$ denotes the hidden $U(1)$ field, $X_{\mu\nu}$ its field strength, and $\chi$
the kinetic mixing parameter. 
We will see in the next subsection that $\chi$ can be tiny.  
Therefore, light hidden photons are also well motivated WISP candidates,

Occasionally, there is also light hidden matter charged under the hidden $U(1)$s.
After diagonalization of the gauge kinetic terms by a shift
$X\to X + \chi A$ and a multiplicative hypercharge renormalization,
one observes that the hidden sector matter particles acquire a minihypercharge, 
$\epsilon = \chi g_h/e$~\cite{Holdom:1985ag}. 
In a similar way minicharged particles can also arise from hidden sector magnetic monopoles
if the gauge fields mix via a non-diagonal $\theta$-term~\cite{Bruemmer:2009ky}.
Therefore, light minicharged particles (MCPs) are another sort of well motivated WISP candidates.

\subsection{Hotspots in axion and other WISPs parameter space from theory}
\label{sec:hotspotstheory}

The masses and couplings of axions and other WISPs to light 
standard model particles appearing in the low energy effective Lagrangians 
(\ref{axion_leff}), (\ref{ALP_leff}), and (\ref{LagKM}) can only be predicted
in terms of more fundamental parameters if an ultraviolet completion of the low energy theory
is specified. 
The most satisfactory ultraviolet completions are arguably the ones 
which are motivated by other issues in particles physics, such as for example  
the unification of fundamental forces, with string theory  
being perhaps the most ambitious
project. 

\begin{figure}[t]
\centerline{\includegraphics[width=0.75\textwidth]{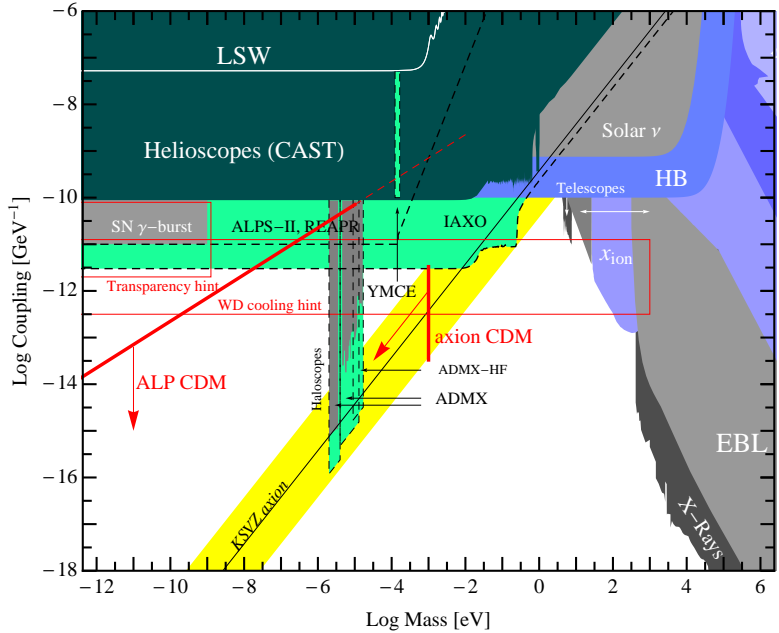}}
\caption{Axion and ALP coupling to photons vs. its mass (adapted from Refs.~\cite{Hewett:2012ns,Baker:2012esg,Cadamuro:2011fd,Arias:2012mb}).
Colored regions are: generic prediction for the QCD axion, exploiting Eqs.~(\ref{axionmass}) and  
(\ref{axionphotoncoupling}), which relate its mass with its coupling to photons (yellow), 
experimentally excluded regions (dark green), constraints from astronomical
observations (gray) or from astrophysical or cosmological arguments (blue), and sensitivity of planned
experiments (light green). Shown in red are boundaries where axions and ALPs can account for all the cold dark matter
produced either thermally or non-thermally by the vacuum-realignment mechanism.
\label{FIG:Roadmap_alp}} 
\end{figure}

\subsubsection{Theory expectations for axions and ALPs}

As already mentioned, in purely field theoretic ultraviolet completions in four dimensions, 
the axion field arises naturally as the phase of a
complex standard model singlet scalar field whose vacuum expectation value $v_{\rm PQ}$ breaks a global 
Peccei-Quinn $U(1)_{\rm PQ}$ symmetry. 
The axion decay constant $f_a$ is then
the symmetry breaking scale of this $U(1)$ symmetry, 
\begin{equation}
f_a = v_{\rm PQ}/C_{ag},
\end{equation} 
and thus still arbitrary. The size of the dimensionless couplings to standard model particles, notably 
the phenomenologically most important ones to photons and electrons, 
\begin{eqnarray}
\mathcal{L} \supset -  \frac{\alpha}{2\pi f_a}  C_{a\gamma}  \,\frac{a}{4} \,F_{\mu\nu} \tilde{F}^{\mu\nu} + \frac{C_{ae}}{2 f_{a}} \,\bar{e} \gamma^\mu\gamma_5 e \partial_\mu a\,,
\label{EQ:DEFCAGG}
\end{eqnarray}
depends on the Peccei-Quinn $U(1)_{\rm PQ}$ charge assignments of standard model matter and 
eventually extra matter beyond the standard model. 
Generically, one expects
\begin{equation}
C_{a\gamma}=\mathcal{O}(1).
\end{equation}
The width of the yellow band in Fig.~\ref{FIG:Roadmap_alp}, showing the generic prediction 
of the coupling $g_{a\gamma}$, Eq.~(\ref{axionphotoncoupling}), of the QCD axion to photons, is supposed to reflect
the possible variations in $C_{a\gamma}$ arising from different field theoretic ultraviolet completions~\cite{Cheng:1995fd}. 
The coupling to electrons, on the other hand, is even more model dependent: it may
be zero at tree level, and only induced at higher orders, leading to $C_{ae}\sim 10^{-4}$. 

Arguably the best motivated purely field theoretic high energy completions yielding a QCD axion 
are the ones in which the global Peccei-Quinn symmetry is not imposed by hand, but in which it appears
as an accidental or automatic consequence of local gauge invariance, renormalisability and the pattern of 
gauge symmetry breakdown~\cite{Georgi:1981pu}. In these cases, the axion decay constant is expected to be of order  
the scale of gauge symmetry breaking, that is, for example, the GUT scale, $f_a\sim v_{\rm GUT}\sim 10^{16}$~GeV, 
in minimal grand unified extensions of the standard model~\cite{Nilles:1981py,Dimopoulos:1982my}, or, an intermediate scale, 
$f_a\sim 10^{10\div 13}$~GeV, in the case of grand unification with a non-minimal particle
content~\cite{Dias:2004hy,Dias:2007vx}.  
Still, in these field theoretic completions there remains the puzzle why the Peccei-Quinn symmetry 
is of such high quality: one has to forbid the occurrence of Planck-suppressed operators violating the 
Peccei-Quinn symmetry with mass dimensions less than ten~\cite{Kamionkowski:1992mf,Barr:1992qq,Holman:1992us},
e.g. by invoking large discrete symmetries, see e.g. Ref.~\cite{Dias:2002gg}. 

The situation may be better in string theoretic ultraviolet completions~\cite{Dine:2010cr}. 
In this case, the axionic shift symmetry of the closed string axions 
arises from the gauge invariance of form fields in ten dimensions. It is valid in all orders of perturbation theory.  
The issue is then whether the closed string axions
remain light in the process of moduli stabilisation~\cite{Banks:2003es,Donoghue:2003vs}. If yes, then the 
QCD axion decay constant  
$f_a$ is expected to be of the order of the string scale $M_s$ ($M_s$ is just the inverse of the fundamental string length).
Typical values of $M_s$, varying between $10^{9}$ and $10^{17}$~GeV, for intermediate scale and GUT scale strings, respectively, suggest values of $f_a$ in the same range. 

\begin{figure}[t]
\centerline{\includegraphics[width=0.9\textwidth]{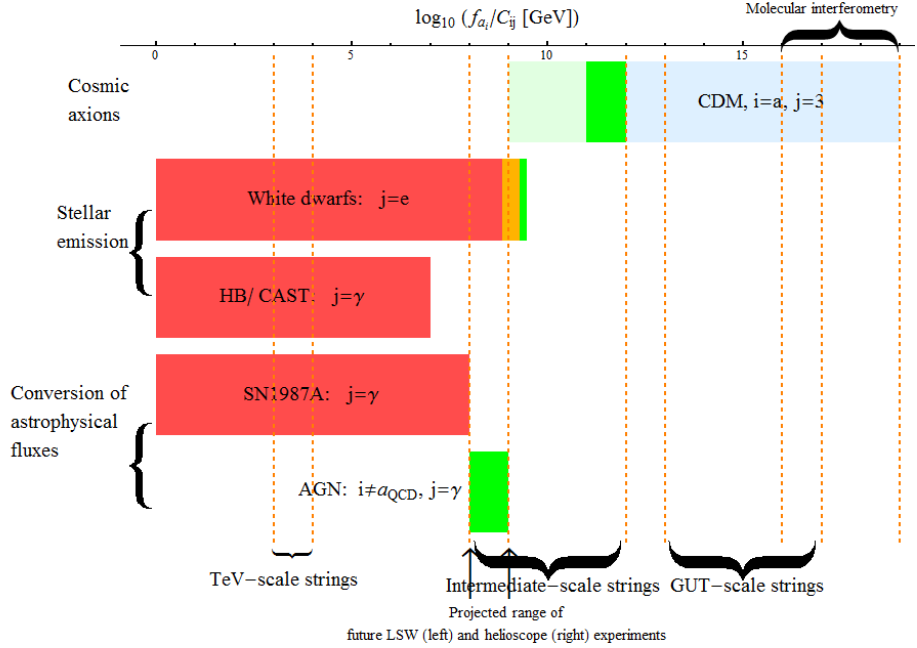}}
\caption{A summary of constraints on and hints for the ratio of decay constants and
couplings, $f_{a_i}/C_{ij}$,  
of the axion and axion-like particles $i$ to fields $j$ (from Ref.~\cite{Cicoli:2012sz}).
The green regions from top to bottom correspond respectively to the classic `axion dark matter window', hints of an axion from
white dwarf cooling and transparency of the Universe to very high energy gamma rays. Red regions are excluded,
and the orange region would be excluded by red giants but is compatible with the hints from white dwarfs. The blue region would be excluded by dark matter overproduction in the absence of a dilution mechanism or tuning of the misalignment angle.
} 
\label{FIG:Constraints}
\end{figure}

Importantly, in the presently best understood string compactification, the 
Large Volume Scenario (LVS) of IIB string theory, there are, after stabilisation of
all moduli, at least two light axions. One of them is associated with the small cycle in Fig.~\ref{Fig:type_ii_comp} 
where the visible sector standard model branes are located: it has 
the properties of the QCD axion. The other one is an ALP: it is associated with the large 
cycle in Fig.~\ref{Fig:type_ii_comp}, much lighter than the QCD axion candidate, and nearly decoupled~\cite{Cicoli:2012sz,Ringwald:2012cu}.  Requiring the gravitino mass 
$m_{3/2}$ and the soft SUSY masses in the TeV regime, the LVS with least fine tuning (tree value of the superpotential 
$W_0\sim 1$) has a string scale of order the intermediate scale\footnote{With some fine-tuning, also an LVS with string scale of order the GUT scale and still TeV scale soft SUSY masses is possible, with correspondingly higher decay constants of order 
$f_a\sim 10^{16}$~GeV.},
\begin{equation}
M_s \sim \frac{M_P}{\sqrt{\mathcal{V}}}\sim  
\sqrt{\frac{M_P\, m_{3/2}}{ W_0}}\sim 10^{9\div 12}\,{\rm GeV}\,,
\label{eq:lvsstringscale}
\end{equation} 
where $\mathcal{V}$ is total volume of the 6D bulk in units of the string length. 
Therefore, it predicts a QCD axion candidate in the 
so-called classic window, 
\begin{equation}
f_a \sim M_s \sim 10^{9\div 12}\,{\rm GeV}\,,
\label{eq:lvsexp}
\end{equation} 
which is phenomenologically
particularly attractive, see Fig.~\ref{FIG:Constraints} and Secs.~\ref{sec:astrophysics} and \ref{sec:cosmology}.  
Generically, these LVS 
compactifications have however several intersecting cycles at the visible sector location and thus 
several 
light closed string ALPs, which have then decay constants
and coupling constants to gauge bosons other than the gluon, e.g. the photon,  
of the same size as the QCD axion candidate, 
\begin{eqnarray}
f_{a_i}&\sim & f_a\sim M_s\sim 10^{9\div 12}\,{\rm GeV},
\label{alpdecayconstant}
\\
C_{i\gamma}& \sim & \mathcal{O}(1)
\Rightarrow  g_{i\gamma} \equiv \frac{\alpha}{2\pi f_{a_i}} C_{i\gamma} 
\sim 10^{-15}\div 10^{-11}\ {\rm GeV}^{-1},
\label{alpcoupling}
\end{eqnarray}
but with a hierarchically smaller mass~\cite{Cicoli:2012sz,Ringwald:2012cu}. The ALPs in the LVS would thus populate
the parameter region to the left of the yellow QCD axion region in Fig.~\ref{FIG:Roadmap_alp}.  
The coupling to electrons is expected to be suppressed by a loop factor,
\begin{equation}
C_{ie}\sim \mathcal{O}(10^{-4}) \,.
\label{alpcouplingelectron}
\end{equation}

A multitude of ALPs may also arise in the context of orbifold compactifications of the
heterotic string as Nambu-Goldstone bosons from the breaking of accidental $U(1)$ symmetries that appear as 
low energy remnants of discrete symmetries~\cite{Choi:2009jt}. In this (model-dependent) case, it is possible to disconnect
the decay constants $f_{a_i}$ from the heterotic string scale, which is of order of the Planck scale,  
and chose them e.g. in the phenomenologically interesting classic axion window.  

Intriguingly, similar models with several ALPs exhibiting an axionic sea-saw pattern of mass scales may, within
a single framework, describe inflation, solve the strong CP problem and  
provide an explanation of cold dark matter and dark energy in the universe~\cite{Chatzistavrakidis:2012bb}.   

\subsubsection{Theory expectation for hidden photons}

Let us now turn to the theoretical expectations of the parameter ranges of hidden photons. 
Kinetic mixing is generated at one-loop by the exchange of heavy messengers that couple both
to the visible $U(1)$ as well as to the hidden $U(1)$. Therefore, in general, one expects  
\begin{equation}
\chi \sim \frac{e g_h}{16\pi^2}\,C,
\end{equation}
where $e$ is the electric charge, $g_h$ is the gauge coupling in the hidden sector, and
$C$ is a dimensionless number. 

In purely field theoretic extensions of the standard model in four dimensions, the hidden sector 
gauge coupling is naturally of order of the visible sector one, $g_h\sim e\sim 0.1$ and $C$ is expected 
to be generically of order one\footnote{It can be much less than
one if the mass splitting $\sim\Delta M$ in the messenger sector is much smaller then the messenger mass $\sim M$, since 
then $C\sim \Delta M/M$~\cite{Dienes:1996zr}.}, leading to the estimate 
\begin{equation}
\chi \sim \frac{e g_h}{16\pi^2}\,C \sim 10^{-4} . 
\end{equation}
 
A similar value is also expected in the case of an ultraviolet completion based on heterotic string theory. 
Indeed, in this case the gauge coupling in the hidden sector is 
predicted to be of the same size as the gauge coupling in the visible sector,  
and the dimensionless constant $C$ is still of order one, as has been demonstrated in explicit 
calculations in heterotic orbifold compactifications~\cite{Goodsell:2011wn}. 
 
Type II string flux compactifications produce a much greater variety of possible values for the mixing $\chi$ 
between brane-localized hidden $U(1)$s and the visible $U(1)$,
cf. Fig.~\ref{Fig:type_ii_comp}. 
This arises mainly from the fact that the hidden sector gauge coupling can be much smaller than the 
visible one, since it is inversely proportional to the extra-dimensional volume $\mathcal{V}_q$ (in units of the 
string length) of the 
$q$-cycle wrapped by the hidden brane: e.g. for a space-time filling $D7$ brane wrapping a 
large four-cycle in an isotropic LVS compactification (see Fig.~\ref{Fig:type_ii_comp}),
\begin{equation}
g_h^2 
\sim \mathcal{V}_4^{-1}
\sim \mathcal{V}^{-2/3}\sim (M_s/M_P)^{-4/3}\sim 10^{-11}
\,,
\label{gsquaredq}
\end{equation}
leading to an estimate 
\begin{equation}
\chi \sim \frac{e}{16\pi^2}\, (M_s/M_P)^{-2/3}\sim 10^{-9}\,,
\end{equation}
for the preferred intermediate string scale $M_s\sim 10^{10}$~GeV. These expectations have been 
verified in explicit calculations~\cite{Abel:2008ai,Goodsell:2009xc}.  
Smaller values of kinetic mixing are obtained in these setups when considering even lower string scales 
by exploiting anisotropic compactifications~\cite{Cicoli:2011yh} or in special cases where the
one-loop contribution is cancelled or vanishes. Moreover,
exponentially suppressed values can be naturally obtained in flux compactifications
with warped throats~\cite{Abel:2008ai}. 

\begin{figure}[t]
\centerline{\includegraphics[width=1.0\textwidth]{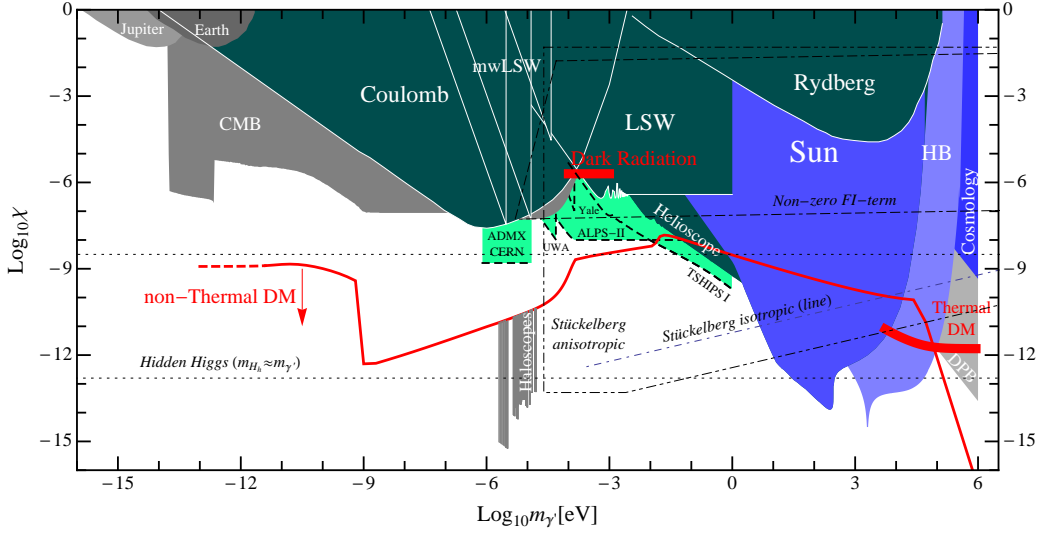}}
\caption{Kinetic mixing parameter vs. hidden photon  mass (adapted from Refs.~\cite{Arias:2012mb,Hewett:2012ns,Baker:2012esg}).
Colored regions are: experimentally excluded regions (dark green), constraints from
astronomical observations (gray) or from astrophysical or cosmological arguments (blue), and sensitivity
of planned experiments (light green). Shown in red are boundaries where the hidden photon would account for all 
cold dark matter produced either thermally or non-thermally by the vacuum-realignment mechanism or where the
hidden photon could account for the hint of dark radiation during the CMB epoch.  
The regions bounded by dotted lines show predictions from 
string theory corresponding to different possibilities for the nature of the hidden photon mass: Hidden-Higgs, a 
Fayet-Iliopoulos term, or the St\"uckelberg mechanism. In general, predictions are uncertain by factors of order one.} 
\label{FIG:Roadmap_hp}
\end{figure}

Masses for the hidden photons can arise via the well-known Higgs mechanism but also via a St\"uckelberg mechanism.
Unfortunately, masses arising from a Higgs mechanism can not be predicted from first principles, 
neither in a purely field theoretic setup nor in a string embedding. They can be tiny, 
if the supersymmetry breaking scale in the hidden sector is much smaller than in the visible sector.

The St\"uckelberg mechanism arises naturally in type II string settings: brane-localised hidden photons 
can get masses by eating the closed string axions associated with the cycles wrapped by the branes 
(cf. Fig.~\ref{Fig:type_ii_comp}). 
In LVS string compactifications, small, even sub-eV, masses can arise, if the 
string scale is sufficiently low~\cite{Goodsell:2009xc,Cicoli:2011yh}.
If the masses arise from a St\"uckelberg mechanism, mass and size of the kinetic mixing are typically linked through one scale, the string scale, and therefore related to each other. Depending on the specific way in which the cycles wrap the singularities one obtains expressions for the masses like
\begin{equation}
m^{\rm st}_{\gamma^{\prime}}\sim 
\frac{M_P}{\mathcal{V}}\sim \frac{M^2_{s}}{M_{P}}\sim 10^2\ {\rm GeV},
\end{equation}
for an intermediate string scale, $M_s\sim 10^{10}$~GeV. A sub-eV St\"uckelberg mass would require
an anisotropic compactification with a string scale in the TeV range. 

Regions in hidden photon parameter space preferred in LVS string compactifications are 
indicated in Fig.~\ref{FIG:Roadmap_hp} by dotted lines  labelled by the different possibilities 
for the origin of the hidden photon mass: a hidden Higgs mechanism, a Fayet-Iliopoulos term, 
or the St\"uckelberg mechanism~\cite{Cicoli:2011yh}. 

\subsection{Constraints on axion and other WISPs parameter spaces from astrophysics and big bang cosmology}
\label{sec:astrophysics}

As we have seen in the last section, the theoretically favored masses and couplings of axions and 
other WISPs span a very wide range in parameter space. Correspondingly, searches for their signatures have 
to exploit a wide variety of observational and experimental techniques, ranging from cosmology and astrophysics
to terrestrial laboratory experiments. The strongest bounds on their existence 
presently often come from their effects on stellar evolution  (cf. Sec.~\ref{sec:stellar}), 
on the propagation of particles through the universe (cf. Sec.~\ref{APP:FLUXES}) and on big bang cosmology
(cf. Sec.~\ref{sec:bigbangcosm}).
However, there are also some intriguing astronomical observations
which are hard to  explain by  known physics and might be interpreted as indirect hints pointing towards
the existence of WISPs.

\subsubsection{Bounds and hints from stellar emission}
\label{sec:stellar}

WISPs, if they exist, are produced in hot astrophysical plasmas such as the interior of stars 
via their coupling to photons or via their coupling to electrons or nucleons.  
They will easily escape, contributing directly to the total energy loss of the star 
and affecting stellar evolution.

For axions and ALPs, one of the strictest bounds on their  
coupling to photons arises from the non-observation of an anomalous energy loss of Horizontal Branch (HB) stars due to 
ALP emission~\cite{Raffelt:1996wa}\footnote{Very recently there has been a new astrophysical constraint 
on the axion-photon coupling based on the non-suppression of the blue loop 
in the evolution of helium-burning massive stars, which  
is mildly more restrictive than the classic HB star limit, 
$g_{i\gamma}\lesssim 8\times 10^{-11}\ {\rm GeV}^{-1}$~\cite{Friedland:2012hj}.}, 
\begin{eqnarray}
g_{i\gamma}\lesssim 10^{-10}\ {\rm GeV}^{-1} \Rightarrow
\frac{f_{a_i}}{C_{i\gamma}}  \gtrsim  10^7\ {\rm GeV},
\label{CASTbound}
\end{eqnarray}
which is shown also in Figs.~\ref{FIG:Roadmap_alp} and  \ref{FIG:Constraints}.  
The best limit on their coupling to electrons arises from white dwarf cooling inferred via pulsationally
unstable white dwarfs (ZZ Ceti stars), where the period decrease
provides a measure of the cooling speed. From the measured period decrease of the star
G117-B15A it was found~\cite{Corsico:2001be,Isern:2003xj,Corsico:2012ki}:
\begin{equation}
g_{ie} < 1.3\times 10^{-13} \Rightarrow \frac{f_{a_i}}{C_{ie}} > 3.9 \times 10^9\ {\rm GeV}.
\end{equation}
Hence, for $C_{ie} \gtrsim 10^{-2}\, C_{i\gamma}$, the limits from electron couplings are more constraining than the 
ones from the two-photon coupling; compare also the excluded red regions labelled as 
``White dwarfs" and ``HB" in Fig.~\ref{FIG:Constraints}. 

However -- and intriguingly -- a reanalysis of the same ZZ Ceti star with improvements in the accuracy and precision of the pulsational data seems to indicate a non-standard energy loss~\cite{Isern:2008nt,Isern:2008fs,Isern:2012ef}. Improvements in the luminosity distribution of white dwarf stars were also found to require additional energy loss, an absence thereof predicting an excess of white dwarfs at luminosities $\log L/L_\odot \simeq -2$. The observed excess in cooling speed is consistent with the existence of an axion or ALP with mass $m_i\lesssim\ {\rm keV}$
and a coupling to electrons of strength:
\begin{equation}
g_{ie} = (2.0\div 7.0)\times 10^{-13}
\Rightarrow \frac{f_{a_i}}{C_{ie}}\simeq
(0.7\div 2.6)\times 10^9\ {\rm GeV}\, .
\label{dec_const_wd}
\end{equation} 

This hotspot in parameter space is shown in green and orange in the line labelled as ``White dwarfs" in Fig.~\ref{FIG:Constraints}. 
Moreover, the red parameter region labelled ``WD cooling hint" in Fig.~\ref{FIG:Roadmap_alp} indicates 
the expectation for the corresponding two-photon coupling, if one allows for a reasonable uncertainty 
in the ratio $C_{i\gamma}/C_{ie}$. 

Clearly, HB stars can also be exploited to bound the parameter space of hidden photons~\cite{Redondo:2008ec}. 
However, in this case also the solar lifetime gives an important and complementary constraint~\cite{Redondo:2008aa}, 
as can be seen by comparing the exclusions regions labelled by ``Sun" and ``HB" in Fig.~\ref{FIG:Roadmap_hp}.

\subsubsection{Bounds and hints from the conversion of astrophysical photon fluxes}
\label{APP:FLUXES}

Due to the anomalous coupling to electromagnetism, $a_i\,F\tilde F\propto a_i\,\vec E\cdot \vec B$, photons 
can oscillate into axions or ALPs in large-scale magnetic fields. These oscillations can
lead to observable effects in astrophysical photon spectra, but with high sensitivity
only for very small masses. Correspondingly, these observations typically constrain only ALPs,
but not the QCD axion.

The absence of a $\gamma$-ray burst in coincidence with neutrinos from the supernova SN 1987A provides, for
$m_i\lesssim 10^{-9}$~eV,  a very restrictive limit on the two-photon coupling of ALPs~\cite{Brockway:1996yr,Grifols:1996id},
\begin{eqnarray}
g_{i \gamma} \lesssim 10^{-11}\, {\rm GeV}^{-1} \Rightarrow
\frac{f_{a_i}}{C_{i\gamma}} 
\gtrsim  10^8\,{\rm GeV},
\end{eqnarray}
an order of magnitude better than the one inferred from the lifetime of HB stars, 
as can be also seen by comparing the ``HB" bound with the ``SN $\gamma$-burst" bound in Fig.~\ref{FIG:Roadmap_alp} 
(see also Fig.~\ref{FIG:Constraints}).

Observations of linearly polarised emission from magnetised
white dwarfs \cite{Gill:2011yp} and changes of the linear polarisation from radio galaxies 
(see, e.g., Ref.~\cite{Horns:2012pp}) provide complementary approaches to search for signatures of ALPs. 
The current limits are close to $g_{i\gamma}\sim 10^{-11}$~GeV$^{-1}$, 
albeit with uncertainties related to the underlying assumptions. 

Surprisingly, the $\gamma$-ray spectra from distant active galactic nuclei (AGN) extend to energies $\gtrsim$~TeV.
This is puzzling,
since the $\gamma$-ray absorption due to $e^+ e^-$ pair production off the extragalactic background light is expected to cut off the spectra at very high energies\footnote{For a recent compilation of the available data and comparison with the expectation,
pointing to a more than four sigma deviation from the expectation, see Ref.~\cite{Horns:2012fx}.}.
Intriguingly, apart from conventional explanations, such a high transparency 
of the Universe may be explained by photon $\leftrightarrow$ ALP oscillations:
the conversion of gamma rays into ALPs in the
magnetic fields around the AGNs or in the intergalactic medium, followed by their unimpeded
travel towards our galaxy and the consequent reconversion into photons in the galactic/intergalactic magnetic
fields. This explanation requires a very light, $m_i\lesssim 10^{-9}~ {\rm eV}$, ALP, which couples to two photons with strength~\cite{DeAngelis:2007dy,Simet:2007sa,SanchezConde:2009wu,Horns:2012kw},
\begin{equation}
g_{i\gamma}\sim  10^{-11} \ {\rm GeV}^{-1} \Rightarrow
f_{a_i}/C_{i\gamma} \sim 10^8\ {\rm GeV}. 
\label{dec_const_transp}
\end{equation}
The corresponding region is shown in red and labelled by ``Transparency hint" in Fig.~\ref{FIG:Roadmap_alp}
(see also the green region labelled ``AGN" in Fig.~\ref{FIG:Constraints}).  
It partly overlaps with the region excluded by the non-observation of a $\gamma$-ray burst from
SN~1987A. 

Remarkably, the values of $f_{a_i}/C_{ie}$ in Eq.~(\ref{dec_const_wd}) and 
$f_{a_i}/C_{i\gamma}$ in Eq.~(\ref{dec_const_transp}), inferred from very
different astronomical observations, turn out to be in the same ballpark.
However, there is one
crucial difference: while the WD value~(\ref{dec_const_wd}) can be realised both with the QCD axion
or with a generic ALP,
the anomalous transparency value~(\ref{dec_const_transp}) can not be realised with the
QCD axion, since the mass of the latter,
$m_a \sim m_\pi f_\pi/f_a\sim 10\ {\rm meV} \left( 10^9\ {\rm GeV}/f_a\right)$,
is much higher than the phenomenologically required values, which are in the sub-neV range.
Intriguingly, the required ALP decay constants, couplings, and mass to explain both anomalies 
are compatible with the generic predictions $f_{a_i}\sim 10^{9\div 12}$~GeV of an intermediate string scale LVS 
compactification with $C_{i\,\gamma}/C_{i\,e} \sim 10$~\cite{Cicoli:2012sz,Ringwald:2012cu}.  

\subsubsection{Bounds and hints from big bang nucleosynthesis and the cosmic microwave background}
\label{sec:bigbangcosm}

Primordial abundances of elements, cosmic microwave background (CMB) anisotropies and large scale structure allow us to infer the  particle content in the early universe. A sensitive measurement of the expansion rate of the universe during big bang nucleosynthesis (BBN) is the $^4$He mass fraction, $Y_p$. 
Several estimates of the helium abundance seem to indicate a (yet not conclusive) excess 
of $Y_p$ which can be attributed to an extra degree of freedom, leading to a higher effective number of  
neutrino species~\cite{Nakayama:2010vs}. Recent determinations of this number seem to favor an excess at the BBN epoch of  $\Delta N_\nu^{\rm eff} \sim 1$ and therefore can be used to motivate or constrain WISPs that may have added new relativistic degrees of freedom. An extra neutral spin-0 particle thermalized 
during BBN is allowed, but it is not the case for a hidden photon (cf. Refs.~\cite{Jaeckel:2010ni,Nakayama:2010vs} and references therein). 

Another sensitive measurement of the number of new light particles is the CMB anisotropy.
Recently, a comprehensive analysis of the effects of a thermal relic population of ALPs with an anomalous coupling to two photons 
on BBN, CMB, and the diffuse photon background establish the most stringent bounds in the mass regime above 
about an eV, see Ref.~\cite{Cadamuro:2011fd} and Fig.~\ref{FIG:Roadmap_alp}. 

Reactions involving $\gamma \rightarrow \rm{WISP}$ conversion would have depleted photons in a frequency-dependent way, that can be tested with the spectrum measurements by FIRAS and have been used to set constraints on light minicharged particles \cite{Melchiorri:2007sq} and hidden photons \cite{Jaeckel:2008fi}. Besides, it has been noted in \cite{Mirizzi:2009iz} that resonant production of hidden photons  would lead to a distortion in the CMB spectrum, providing a strong constraint on these particles, see the excluded region labelled ``CMB" in 
Fig.~\ref{FIG:Roadmap_hp}.

Resonant oscillations between photons and hidden photons after BBN and before CMB decoupling may comprise a hidden CMB. 
The hidden photons in the range of $m_{{\gamma'}}\sim$~meV and $\chi \sim 10^{{-6}}$  produced by such oscillations  would constitute a contribution to the dark radiation at the CMB epoch, leading to an
apparent increase of the effective number of neutrinos~\cite{Jaeckel:2008fi}\footnote{Recently, it was pointed out that
the dark radiation may also be comprised by the ALP associated with the large cycle in a sequestered LVS with 
string scale of order $10^{13}$~GeV~\cite{Cicoli:2012aq,Higaki:2012ar}.}. 
Interestingly, according to several recent observations, a number higher than three at the CMB epoch is currently favored \cite{Komatsu:2010fb,Dunkley:2010ge}. 
This observation will soon be tested by the measurements of the CMB anisotropy by the PLANCK satellite. 
The parameter region required to explain the 
current excess by hidden photons is labelled in red by ``Dark Radiation" in Fig.~\ref{FIG:Roadmap_hp}.

\section{WISPy cold dark matter}
\label{sec:cosmology}

QCD axions are very well motivated candidates for cold dark matter~\cite{Preskill:1982cy,Abbott:1982af,Dine:1982ah}. 
Axion cold dark matter is produced non-thermally in the early universe by the vacuum-realignment
mechanism and, in some models and under certain circumstances, 
also via the decay of topological defects such as axion strings and domain walls 
(for a review, see Ref.~\cite{Sikivie:2006ni}). 
Importantly, the vacuum-realignment mechanism  does not work exclusively for QCD axions. It works in fact  
generically for bosonic WISPs such as ALPs~\cite{Arvanitaki:2009fg,Acharya:2010zx,Arias:2012mb} and hidden photons~\cite{Arias:2012mb,Nelson:2011sf}. 
Once produced, a population of very light cold dark matter particles is extremely difficult to reabsorb by the primordial plasma. Therefore, for ALPs and hidden photons, 
a huge region in parameter space spanned by their masses and their couplings to standard model particles 
can give rise to the observed dark matter.

The vacuum-realignment mechanism relies on assuming that fields in the early universe have a random initial state
(arising from quantum fluctuations during inflation) which is fixed by the cosmic expansion. Fields with mass $m_i$ evolve on timescales $t \sim m_i^{-1}$. After such a timescale, the fields respond by attempting to minimise their potential, and consequently oscillate around the minimum. If there is no significant damping via decays, these oscillations can behave as a cold dark matter fluid since their energy density is diluted by the cosmic expansion as $\rho_i\propto a^{-3}$, where $a$ is the 
cosmic scale factor. 

Assuming that the reheating temperature after inflation
is below $f_a$ and that there is no dilution by, e.g., late decays of particles beyond the standard model
(e.g. moduli), the expected cosmic mass fraction in QCD axion cold dark matter is then~\cite{Sikivie:2006ni} 
\begin{equation}
\Omega_a h^2\approx  0.71 \times \left( \frac{f_a}{10^{12}\ \rm GeV} \right)^{7/6} \left( \frac{\Theta_a}{\pi} \right)^2,
\label{eq:omegaqcdaxion}
\end{equation}
where $\Theta_a$ is the initial misalignment angle, 
 while the one in ALP cold dark matter is~\cite{Arias:2012mb} 
\begin{equation}
\Omega_{a_i} h^2 \approx 0.16 \times
\left( \frac{m_i}{\rm eV} \right)^{1/2} \left( \frac{f_{a_i}}{10^{11}\ \rm GeV} \right)^{2}  \left( \frac{\Theta_i}{\pi} \right)^2,
\label{eq:omegaalp}
\end{equation}
where $m_i$ is the mass of the ALP. 
Therefore, the QCD axion can be the dominant part 
of cold dark matter if its decay constant exceeds $f_a\gtrsim 10^{11}$~GeV, corresponding to a mass 
$m_a\lesssim 10^{-4}$~eV. That is, the QCD axion predicted, e.g., in an intermediate string scale LVS 
may qualify for a good dark matter candidate~\cite{Cicoli:2012sz,Ringwald:2012cu}. 
A QCD axion with a GUT scale decay constant, $f_a\sim 10^{16}$~GeV, on the other hand, would
overclose the universe, unless the initial misaligment angle is very small, $\Theta_a\sim 10^{-3}$ 
(``anthropic axion window"~\cite{Tegmark:2005dy,Mack:2009hv}), or there is a late dilution.
Another consequence of the assumption that the reheat temperature after inflation is below $f_a$ 
is the fact that, in this case, the axion field is present during inflation.
Correspondingly, it is subject to quantum fluctuations, leading to isocurvature
fluctuations that are severely constrained~\cite{Beltran:2006sq,Hertzberg:2008wr,Hamann:2009yf}.

Conversely, if the reheating temperature after inflation is above $f_a$, 
the initial misalignment angles take on different values in different patches of the universe,
leading to an average contribution~\cite{Sikivie:2006ni}   
\begin{equation}
\Omega_a h^2\approx  0.3 \times \left( \frac{f_a}{10^{12}\ \rm GeV} \right)^{7/6} 
\label{eq:omegaqcdaxionav}
\end{equation}
to the cosmic mass fraction. Therefore, in this case, $m_a \sim 10^{-5}$~eV axions 
can provide dark matter, but smaller masses are excluded (in the absence of a dilution mechanism). 
In this case, however, the decay of cosmic strings and domain walls may provide for additional
sources for non-thermal axions~\cite{Sikivie:2006ni}. 
These extend the region where axions may provide cold dark matter to
larger values of $m_a\sim 10^{-3}$~eV~\cite{Wantz:2009it,Hiramatsu:2010yu,Hiramatsu:2012gg}.
Moreover, the spatial axion
density variations are large at the QCD transition and they are
not erased by free streaming. When matter begins to dominate
the universe, gravitationally bound ``axion mini clusters" form
promptly~\cite{Kolb:1993zz,Zurek:2006sy}. A significant fraction of cold dark matter axions can reside
in these bound objects.

The values of the decay constant where the QCD axion can be the dominant part of 
dark matter are indicated in Fig.~\ref{FIG:Constraints}: the green region corresponds 
to the ``classic axion cold dark matter window", $f_a\sim 10^{11\div 12}$~GeV, 
the blue region is the ``anthropic" region, $f_a \gtrsim 10^{12}$~GeV, while for the light green region,  
$f_a\sim 10^{9\div 11}$~GeV, one has to envoke non-thermal production mechanisms 
beyond vacuum-realignment such as decaying cosmic strings and domain walls. 
The corresponding region in the photon coupling of the axion is indicated in red in 
Fig.~\ref{FIG:Roadmap_alp} and labelled by ``axion CDM". In the same figure, 
we display also the region where ALPs produced by the vacuum-realignment mechanism 
can be the cold dark matter (labelled in red by ``ALP CDM"). Here, in generalisation
to Eq.~(\ref{eq:omegaalp}), a general variation of ALP mass with temperature 
has also been allowed~\cite{Arias:2012mb}.   

In supersymmetric models, the computation of the dark matter abundance becomes 
more complicated in that now there is the possibility of mixed axion-LSP
cold dark matter, {\it i.e.} two co-existing dark matter particles.
In the case where the LSP is the lightest neutralino, then neutralinos 
can be produced thermally as usual, but they can also be produced via 
thermal production of axinos which are produced 
after the end of inflation. The temperature at which axinos decay is typically
at the tens of MeV scale, and if axinos temporarily dominate the energy 
density of the universe, then neutralinos may experience a period of 
re-annihilation after axino decay leading to an increased abundance~\cite{Choi:2008zq}.

Also, saxions may be produced either thermally or via the vacuum realignment mechanism 
and may also temporarily dominate the energy density of the universe.
They may decay into $aa$, $gg$ and perhaps $\tilde{g}\tilde{g}$ or
$\tilde{a}\tilde{a}$. The first of these leads to dark radiation. 
The second decay mode, if dominant, may lead to large late-time 
(temperatures of tens of MeV) entropy injection, leading to possible dilution
of all relics present, including LSPs, axions and baryons. 
The third and fourth decays may again augment the LSP abundance.

A complete calculation requires eight or nine coupled Boltzmann equations
which describe the relative abundances of radiation, neutralinos, gravitinos,
axinos and separately thermally produced axions and saxions, and  
axions and saxions produced via the vacuum realignment mechanism. 
The ninth equation would track relativistic axions produced from saxion decay.
Such calculations have been carried out in Refs.~\cite{Baer:2011hx,Baer:2011uz}  
and further work is ongoing. 
The upshot of these analyses is that the neutralino abundance almost 
always gets increased beyond its thermal-production-only value, which
then would favor models with a typical neutralino underabundance which
occurs for higgsino-like or wino-like neutralinos. For large values of $f_a$, 
then saxions from the vacuum-realignment mechanism may produce large relic dilution via entropy dumping,
thus allowing for much larger values of $f_a$ sometimes as high as 
approaching the GUT scale.
Depending on the values of the parameters $m_{\tilde{a}}$, $m_s$, 
$f_a$, $m_{3/2}$, the initial field values $\Theta_a$ and $s(t_i)$ and 
the entire SUSY particle spectrum and the particular SUSY axion model, 
then the dark matter may be either
neutralino- or axion-dominated, or a comparable mixture. In such scenarios,
one might expect eventual direct detection of both relic WIMPs and relic axions.

The vacuum-realignment mechanism can also be at work to generate a population of hidden photons 
which may survive as a significant contribution to cold dark matter today~\cite{Nelson:2011sf,Arias:2012mb}. 
Taking into account various constraints,  arising, e.g., from the possible decay of the
hidden photon condensate and from the non-observation of CMB distortions due to hidden photon--photon 
oscillations,  
still a huge region in hidden photon parameter space can explain the observed dark matter, as
can be seen by the red region labelled by ``non-Thermal DM" in Fig.~\ref{FIG:Roadmap_hp}. 
This region has partial overlap with the regions motivated by theory (dotted lines).

To conclude this section let us note that dark matter generated by the vacuum-realignment mechanism may have interesting properties beyond those of cold dark matter. At the time of their production, particles produced by this mechanism are semi-relativistic: their momenta are of the order of the Hubble constant $p\sim H\sim m$ and their velocity distribution has, accordingly,
a very narrow width. Combined with the high number density of particles, 
this narrow distribution typically leads to very high occupation numbers for each quantum state~\cite{Arias:2012mb},
\begin{equation}
{\cal N}_{\rm occupation}
\sim 10^{42} \left(\frac{\rm eV}{m}\right)^{5/2}.
\end{equation}
If the interactions are strong enough to achieve thermalisation, as argued in Refs.~\cite{Sikivie:2009qn,Erken:2011dz} for the case of axions, this high occupation number can lead to the formation of a Bose-Einstein condensate. This could lead to interesting properties which may also lead to peculiar signatures in cosmological observations, such as characteristic 
structures in galactic dark matter halos~\cite{Sikivie:2009qn,Erken:2011dz,Sikivie:2010yn,Erken:2011xj,Kain:2011pd},
which may have even been observed already~\cite{Sikivie:2012gi}.  
Clearly, these features could also be realised for other WISPs produced by the vacuum-realignment mechanism, 
such as ALPs and hidden photons.

\section{Experimental searches for axions and other WISPs}
\label{sec:searches}

Experimental searches for WISPs require different approaches to those usually employed
in particle physics. In particular, the very weak interactions involved demand experiments at
the low-energy, high-intensity frontier~\cite{Jaeckel:2010ni,Hewett:2012ns,Baker:2012esg}. 
At low energies, high-power cw or high-intensity pulsed lasers~\cite{Dobrich:2010hi,Dobrich:2010ie}, 
strong magnetic fields, radio
frequency technology as well as other techniques allow us to explore couplings many orders of
magnitude smaller than those probed in collider experiments at the high-energy frontier. 
A significant community has grown
over the last ten years and several experiments have already demonstrated the power of these
approaches in relatively modest setups. Compared to collider experiments, this is still a very
young field. Existing experiments can still be scaled up to achieve sensitivity gains of many
orders of magnitude and new experimental ideas arise very frequently with the potential of
complementing existing schemes. 

\subsection{Haloscope searches}

For decay constants in the range $f_a\gtrsim 10^{11\div 12}$~GeV, the QCD axion is expected to 
contribute substantially to the cold dark matter in the universe, see Eq.~(\ref{eq:omegaqcdaxion}).  
Galactic halo QCD axions, as well as ALPs -- if they couple to photons,
$C_{i\gamma}\neq 0$ -- can then be directly searched for~\cite{Sikivie:1983ip} via their resonant conversion 
in a high-Q electromagnetic cavity permeated by a strong static magnetic field into
a quasi-monochromatic microwave signal with frequency 
$$\nu=m_a/(2\pi )=0.24\ {\rm GHz}\times (m_a/\mu{\rm eV}).$$
A number of experiments of this type have already been done~\cite{DePanfilis:1987dk,Wuensch:1989sa,Hagmann:1990tj,Asztalos:2001tf,Asztalos:2009yp} and further improvements are
underway~\cite{Asztalos:2011bm,Heilman:2010zz}.
These haloscope searches reach currently sensitivities of order
\begin{equation}
g_{a\gamma}\sim g_{i\gamma}\sim 10^{-(15\,\div 13)}\ {\rm GeV}^{-1}
\Rightarrow
\frac{f_{a_i}}{C_{i\gamma}}\gtrsim 10^{10\,\div 12}\ {\rm GeV},
\end{equation}
in the $(1\div 10)\ \mu$eV mass range, as can be seen from the gray 
(established limits) and green (projected sensitivities) areas in Fig.~\ref{FIG:Roadmap_alp}
labelled as ``Haloscopes". A substantial range of the ``classic QCD axion dark matter" parameter range 
(and similarly ALP cold dark matter range) 
will be probed by the Axion Dark Matter eXperiment ADMX, ADMX-HF, and perhaps new haloscopes currently under investigation, 
which aim to exploit available microwave cavities and 
magnets~\cite{Baker:2011na}, such as the Tore Supra magnet in Cadarache or the H1 solenoid at the 
decommissioned HERA accelerator at DESY. The Yale Microwawe Cavity Experiment (YMCE) will probe 
ALP dark matter with a photon coupling of order $10^{-10}$~GeV$^{-1}$ and a mass of order 10 meV 
(cf. Fig.~\ref{FIG:Roadmap_alp}).    

A possible new haloscope concept based on axion absorption and emission
between electromagnetic fields within a high finesse optical
cavity has recently been proposed in Ref.~\cite{Melissinos:2008vn}. Here, the axion cold dark matter signal 
appears as resonant sidebands on the carrier. This technique should be sensitive in the axion
or ALP mass range $m_{a_i}\sim 10^{-6}\div 10^{-4}$~eV.  

While the above haloscope search techniques are restricted to the mass region $m_{a_i}\gtrsim 10^{-7}$~eV 
and thus, in the case of the QCD axion, to decay constants less than $f_a\lesssim 10^{13}$~GeV, 
another possibility to search directly for QCD axion cold dark matter most sensitive in 
the range  $f_a\sim 10^{16\div 19}$~GeV has recently been pointed out in  
Ref.~\cite{Graham:2011qk}: the idea is to search for oscillating nuclear electric dipole moments 
induced by the oscillating galactic dark matter axion field. Conceivably,
these oscillating electric dipole moments could be measured by 
extremely tiny energy shifts in cold molecules, exploiting molecular interferometry.

Microwave cavity experiments looking for cold dark matter axions can also be used to constrain and search for galactic
hidden photon dark matter~\cite{Arias:2012mb}. In the mass range 
$m_{\gamma^\prime}\sim 10^{-6}\div 10^{-5}$~eV, the current bounds on 
hidden photon dark matter are already quite impressive, as apparent from the grey region
labelled as ``Haloscopes" in Fig.~\ref{FIG:Roadmap_hp}. At smaller masses, in particular
below $10^{-9}$~eV, one may try to look for hidden photon (and perhaps also 
ALP) dark matter using antennas rather than cavities, 
having admittedly lower sensitivity, but over a broad range of masses~\cite{Horns:inprep}.   

\subsection{Helioscope searches}

The sun is potentially a strong source of different kinds of WISPs. In fact,
the flux of WISPs from the sun could reach up to 10\% of the solar photon luminosity without
being in conflict with helioseismological and solar neutrino data.

With axion helioscopes~\cite{Sikivie:1983ip} one tries to detect
solar axions by their conversion into photons inside of a strong magnet pointed towards the sun. 
The currently best sensitivity for solar QCD axions and ALPs has been obtained
by the CERN Axion Solar Telescope (CAST), employing an LHC dipole test magnet with an elaborate 
elevation and azimuth drive to track the sun. As can be seen in Fig.~\ref{FIG:Roadmap_alp}, 
the CAST limits (labelled ``Helioscopes (CAST)") have already surpassed, at low masses, 
the bounds from HB stars~\cite{Andriamonje:2007ew}, thus exploring previously uncharted ALP parameter space. 
At larger masses, CAST has started to probe the predictions for 
QCD axions (yellow band)~\cite{Arik:2008mq,Arik:2011rx}.  
A proposed fourth-generation axion helioscope~\cite{Irastorza:2011gs}, dubbed the International Axion Observatory (IAXO),
envisions a huge toroidal magnet especially designed for the experiment, 
a detection system consisting of large X-ray telescopes coupled to
ultra-low background X-ray detectors, and a large, robust tracking system. IAXO would enable
the search for solar axions and ALPs in a broad mass range, as shown by the green region 
labelled ``IAXO" in Fig.~\ref{FIG:Roadmap_alp}, allowing to probe a part of QCD axion parameter space, 
corresponding to $f_a\sim 10^{9\div 10}$~GeV, which will not be covered by the haloscope searches.     

A very welcome feature of helioscope searches for axions and ALPs is that they are sensitive down to 
arbitrarily low masses: in fact,  
their projected sensitivity is best and stays constant at small masses, see Fig.~\ref{FIG:Roadmap_alp}. 
That means, with IAXO one may probe the intermediate string scale LVS, in particular the possible existence of more 
ALPs with approximately the same coupling to photons as the QCD axion, testing also the ALP explanation
of the puzzles from astrophysics, such as the anomalous transparency of the universe for very high 
energy photons and the anomalous energy loss of white dwarfs. In fact,  
the projected sensitivity of IAXO nicely overlaps with the ALP parameter region required to
explain these puzzles, see Fig.~\ref{FIG:Roadmap_alp}.

Helioscopes can also be used to search for solar hidden photons, by trying to detect
them by their oscillations into photons inside the vacuum chamber pointing to the sun~\cite{Redondo:2008aa}.
The axion and ALP limits from CAST, translated into hidden photons, are displayed in 
Fig.~\ref{FIG:Roadmap_hp} and labelled as ``Helioscope". The Solar Hidden Photon Search 
(SHIPS) in Hamburg is a dedicated helioscope presently being commissioned~\cite{Schwarz:2011gu}, 
exploring previously uncharted regions of hidden photon parameter space in the 
mass range $m_{\gamma^\prime}\sim 10^{-3}\div 1$~eV (labelled ``TSHIPS" in Fig.~\ref{FIG:Roadmap_hp}), 
in particular regions motivated by dark radiation and dark matter.  

Solar ALPs can be efficiently converted into photons in the high purity low background crystals used in some
WIMP dark matter experiments or in searches for neutrino-less double-beta decay, but 
presently the inferrred bounds are not competitive to the bounds from CAST at low masses. 
 
\subsection{Light-shining-through-walls searches} 

Intriguingly, the ALPs parameter region required for the explanation of the above mentioned astro puzzles 
can also be partially probed by purely laboratory based 
light-shining-through-walls (LSW) experiments~\cite{Anselm:1986gz,VanBibber:1987rq,Redondo:2010dp}, 
where laser photons are sent along 
a strong magnetic field, allowing for their conversion into ALPs, which may then reconvert  
in the strong magnetic field behind a blocking wall into photons, apparently shining through the
wall and susceptible to detection. For hidden photon searches, the magnetic field is not 
needed, since photons can oscillate into hidden photons already in field-free 
vacuum~\cite{Okun:1982xi,Ahlers:2007rd,Ahlers:2007qf}. 
For minicharged particles, specialised magnetic field configurations in LSW setups have been shown to be 
favourable~\cite{Dobrich:2012sw,Dobrich:2012jd}. 
Current LSW experiments are not yet as sensitive as halo- and
helioscopes, at least for axions and ALPs\footnote{For hidden photons, LSW experiments are 
superior to helioscopes in certain mass ranges, cf. Fig.~\ref{FIG:Roadmap_hp}.}, but offer full control over the WISP production and do not rely on astrophysical or
cosmological assumptions~\cite{Jaeckel:2006xm}.

The currently best WISP sensitivity exploiting the LSW technique 
has been reached by the first Any Light Particle Search (ALPS-I) 
experiment in Hamburg~\cite{Ehret:2009sq,Ehret:2010mh}, which was set
up using a superconducting HERA dipole magnet at the site of DESY, with the first half of the magnet,
the production region, being enclosed in an optical resonator (to amplify photon-WISP conversions) 
and the second half being used as the regeneration region. The end of the
regeneration tube was connected to a light-tight box in which signal photons were redirected to
a CCD camera. The bounds established by ALPS-I for ALPs and hidden photons are labelled by ``LSW"  
in Figs.~\ref{FIG:Roadmap_alp} and \ref{FIG:Roadmap_hp}, respectively. 
For the ALPs case, they are not competitive with the astro bounds from HB stars,
but for the hidden photon case they are, in the meV mass region, the most stringent
ones and exclude already a sizeable region suggested by a possible explanation for the dark
radiation in the universe. 

There are two proposals for the next-generation of LSW experiments: ALPS-II~\cite{alpsiitdr} at DESY and
the Resonantly Enhanced Axion-Photon Regeneration Experiment (REAPR) at Fermilab,
aiming at surpassing the current helioscope bounds from CAST and tackling some of the ALP 
parameter space favored by astrophysical observations, cf. the light-green region in 
Fig.~\ref{FIG:Roadmap_alp} labelled by ``ALPS-II, REAPR". 
ALPS-II proposes to use 10+10 straightened HERA magnets~\cite{Ringwald:2003nsa}, a high-power laser
system, a superconducting low-background detector and the pioneering realization of an optical
regeneration cavity~\cite{Hoogeveen:1990vq}. 
REAPR is proposing a setup using Tevatron dipoles and a different scheme to lock the optical 
cavities~\cite{Mueller:2009wt}.

LSW experiments can also be performed in other spectral ranges, notably in the 
microwave~\cite{Hoogeveen:1992nq,Jaeckel:2007ch,Caspers:2009cj} 
and in the X-ray ranges~\cite{Rabadan:2005dm,Dias:2009ph}. 
Currently, the microwave LSW technique is most promising for hidden photons, 
as can be seen from the current bounds labelled as 
``mwLSW"~\cite{Wagner:2010mi,Betz:2012tp} and the projected sensitivities labelled as ``ADMX, CERN" in 
Fig.~\ref{FIG:Roadmap_hp} (for a pionieering X-ray LSW experiment, see Ref.~\cite{Battesti:2010dm}). 

\section{Conclusions}
\label{sec:conclusions}

As the experiments at the Large Hadron Collider have not yet observed any hint of 
a WIMP and after a decade of enormous
progress in WIMP direct detection -- more than 3 orders of magnitude in sensitivity in the 
WIMP-nucleon coupling constant -- without a clear positive signal, 
the axion dark matter hypothesis stands out as increasingly interesting and deserves serious attention. 
The cosmological implications of
the axion are well founded and represent a powerful motivation for an enhanced experimental
program. 

Along this way there are many opportunities to detect other WISPs, notably 
axion-like particles and hidden photons, which are also strongly motivated from embeddings 
of the standard model into more fundamental theories. 
In fact, a number of small-scale experiments at the low-energy, high-intensity frontier  -- haloscopes, helioscopes, 
and light-shining-through-a-wall -- are actively searching for these elusive particles, complementing
searches for physics beyond the standard model at the high-energy frontier. A plausible next
generation of experiments includes scaled-up versions of the existing techniques as
well as innovative concepts, together covering a huge unexplored parameter space.
A WISP discovery would have a tremendous impact on our understanding of 
fundamental physics, astrophysics and may shed light on the dark universe.

\section*{Acknowledgments}

We warmly thank Howard Baer, Michele Cicoli, 
Michael Dine, Babette D\"obrich, Mark Goodsell, 
Joerg Jaeckel, Axel Lindner, Hans Peter Nilles, Roberto Peccei, Georg Raffelt, Javier Redondo, and Leslie Rosenberg  
for many discussions, close collaboration, and feedback on the manuscript.






\begin{thebibliography}{00}



\bibitem{Jaeckel:2010ni}
  J.~Jaeckel, A.~Ringwald,
  The Low-Energy Frontier of Particle Physics, 
  Ann.\ Rev.\ Nucl.\ Part.\ Sci.  { 60} (2010) 405
  [arXiv:1002.0329 [hep-ph]].

\bibitem{Hewett:2012ns}
  J.~L.~Hewett {\it et al.}, 
  Fundamental Physics at the Intensity Frontier, 
  arXiv:1205.2671 [hep-ex].
  
\bibitem{Baker:2012esg}
  O.~K.~Baker {\it et al.},
  Fundamental Physics at Low Energies - The Quest for Axions and Other New Light Particles, 
  Input to the European Strategy on Particle Physics 2012.
  
\bibitem{Beringer:1900zz}
  J.~Beringer {\it et al.}  [Particle Data Group Collaboration],
  Review of Particle Physics (RPP), 
  Phys.\ Rev.\ D {86} (2012) 010001.

\bibitem{Peccei:1977hh}
  R.~D.~Peccei, H.~R.~Quinn,
  CP Conservation in the Presence of Instantons,
  Phys.\ Rev.\ Lett.\  {38} (1977) 1440.

\bibitem{Georgi:1986df}
  H.~Georgi, D.~B.~Kaplan, L.~Randall,
  Manifesting The Invisible Axion At Low-energies,
  Phys.\ Lett.\ B {169} (1986) 73.

\bibitem{Weinberg:1977ma}
  S.~Weinberg,
  A New Light Boson?,
  Phys.\ Rev.\ Lett.\  {40} (1978) 223.
  
\bibitem{Wilczek:1977pj}
  F.~Wilczek,
  Problem of Strong P and T Invariance in the Presence of Instantons,
  Phys.\ Rev.\ Lett.\  {40} (1978) 279.

\bibitem{Kim:1979if} 
  J.~E.~Kim,
  Weak Interaction Singlet and Strong CP Invariance,
  Phys.\ Rev.\ Lett.\  {43} (1979) 103.
  
\bibitem{Dine:1981rt} 
  M.~Dine, W.~Fischler, M.~Srednicki,
  A Simple Solution to the Strong CP Problem with a Harmless Axion,
  Phys.\ Lett.\ B {104} (1981) 199.
  
\bibitem{Shifman:1979if} 
  M.~A.~Shifman, A.~I.~Vainshtein, V.~I.~Zakharov,
  Can Confinement Ensure Natural CP Invariance of Strong Interactions?,
  Nucl.\ Phys.\ B {166} (1980) 493.
  
\bibitem{Zhitnitsky:1980tq} 
  A.~R.~Zhitnitsky,
  On Possible Suppression of the Axion Hadron Interactions, (In Russian),
  Sov.\ J.\ Nucl.\ Phys.\  { 31} (1980) 260
  [Yad.\ Fiz.\  {31} (1980) 497].
  
        
\bibitem{Bardeen:1977bd} 
  W.~A.~Bardeen, S.~-H.~H.~Tye,
  Current Algebra Applied to Properties of the Light Higgs Boson,
  Phys.\ Lett.\ B {74} (1978) 229.
  
\bibitem{Kaplan:1985dv} 
  D.~B.~Kaplan,
  Opening the Axion Window,
  Nucl.\ Phys.\ B {260} (1985) 215.
  
\bibitem{Srednicki:1985xd} 
  M.~Srednicki,
  Axion Couplings to Matter. 1. CP Conserving Parts,
  Nucl.\ Phys.\ B {260} (1985) 689.

\bibitem{Witten:1984dg} 
  E.~Witten,
  Some Properties of O(32) Superstrings,
  Phys.\ Lett.\ B {149}, 351 (1984) 351.
  
\bibitem{Conlon:2006tq} 
  J.~P.~Conlon,
  The QCD Axion and Moduli Stabilisation,
  JHEP {0605} (2006) 078
  [hep-th/0602233].

\bibitem{Svrcek:2006yi} 
  P.~Svrcek, E.~Witten,
  Axions in String Theory,
  JHEP {0606} (2006) 051
  [hep-th/0605206].

\bibitem{Arvanitaki:2009fg} 
  A.~Arvanitaki, S.~Dimopoulos, S.~Dubovsky, N.~Kaloper, J.~March-Russell,
  String Axiverse,
  Phys.\ Rev.\ D {81} (2010) 123530
  [arXiv:0905.4720 [hep-th]].

\bibitem{Acharya:2010zx} 
  B.~S.~Acharya, K.~Bobkov, P.~Kumar,
  An M Theory Solution to the Strong CP Problem and Constraints on the Axiverse,
  JHEP {1011} (2010) 105
  [arXiv:1004.5138 [hep-th]].

\bibitem{Cicoli:2012sz} 
  M.~Cicoli, M.~Goodsell, A.~Ringwald,
  The Type IIB String Axiverse and its Low-Energy Phenomenology,
  arXiv:1206.0819 [hep-th].

\bibitem{Ringwald:2012cu}
  A.~Ringwald,
  Searching for Axions and ALPs from String Theory,
  arXiv:1209.2299 [hep-ph].


\bibitem{Goodsell:2010ie}
  M.~Goodsell, A.~Ringwald,
  Light Hidden-Sector U(1)s in String Compactifications, 
  Fortsch.\ Phys.\  {58} (2010) 716
  [arXiv:1002.1840 [hep-th]].

\bibitem{Holdom:1985ag}
  B.~Holdom,
  Two U(1)'s and Epsilon Charge Shifts,
  Phys.\ Lett.\ B {166} (1986) 196.

\bibitem{Bruemmer:2009ky}
  F.~Brummer, J.~Jaeckel, V.~V.~Khoze,
  Magnetic Mixing: Electric Minicharges from Magnetic Monopoles,
  JHEP {0906} (2009) 037
  [arXiv:0905.0633 [hep-ph]].

\bibitem{Cadamuro:2011fd}
  D.~Cadamuro, J.~Redondo,
  Cosmological Bounds on Pseudo Nambu-Goldstone Bosons,
  JCAP {1202} (2012) 032
  [arXiv:1110.2895 [hep-ph]].

\bibitem{Arias:2012mb}
  P.~Arias, D.~Cadamuro, M.~Goodsell, J.~Jaeckel, J.~Redondo, A.~Ringwald,
  WISPy Cold Dark Matter,
  JCAP {1206} (2012) 013
  [arXiv:1201.5902 [hep-ph]].


\bibitem{Cheng:1995fd}
  S.~L.~Cheng, C.~Q.~Geng, W.~T.~Ni,
  Axion - Photon Couplings in Invisible Axion Models,
  Phys.\ Rev.\ D {52} (1995) 3132
  [hep-ph/9506295].

\bibitem{Georgi:1981pu}
  H.~M.~Georgi, L.~J.~Hall, M.~B.~Wise,
  Grand Unified Models With An Automatic Peccei-quinn Symmetry,
  Nucl.\ Phys.\ B {192} (1981) 409.

\bibitem{Nilles:1981py}
  H.~P.~Nilles, S.~Raby,
  Supersymmetry and the Strong CP Problem,
  Nucl.\ Phys.\ B {198} (1982) 102.
  
\bibitem{Dimopoulos:1982my}
  S.~Dimopoulos, P.~H.~Frampton, H.~Georgi, M.~B.~Wise,
  Automatic Invisible Axion Without Domain Walls,
  Phys.\ Lett.\ B {117} (1982) 185.

\bibitem{Dias:2004hy}
  A.~G.~Dias, V.~Pleitez,
  Grand Unification and Proton Stability near the Peccei-Quinn Scale,
  Phys.\ Rev.\ D {70} (2004) 055009
  [hep-ph/0407074].
  
\bibitem{Dias:2007vx}
  A.~G.~Dias, E.~T.~Franco, V.~Pleitez,
  An SU(5) x Z(13) Grand Unification Model,
  Phys.\ Rev.\ D {76} (2007) 115010
  [arXiv:0708.1009 [hep-ph]].
    
  

\bibitem{Kamionkowski:1992mf}
  M.~Kamionkowski, J.~March-Russell,
  Planck Scale Physics and the Peccei-Quinn Mechanism,
  Phys.\ Lett.\ B {282} (1992) 137
  [hep-th/9202003].

\bibitem{Barr:1992qq}
  S.~M.~Barr, D.~Seckel,
  Planck Scale Corrections to Axion Models,
  Phys.\ Rev.\ D {46} (1992) 539.
  
\bibitem{Holman:1992us}
  R.~Holman, S.~D.~H.~Hsu, T.~W.~Kephart, E.~W.~Kolb, R.~Watkins, L.~M.~Widrow,
  Solutions to the Strong CP Problem in a World with Gravity,
  Phys.\ Lett.\ B {282} (1992) 132
  [hep-ph/9203206].
  
\bibitem{Dias:2002gg}
  A.~G.~Dias, V.~Pleitez, M.~D.~Tonasse,
  Naturally Light Invisible Axion in Models with Large Local Discrete Symmetries,
  Phys.\ Rev.\ D {67} (2003) 095008
  [hep-ph/0211107].
  
\bibitem{Dine:2010cr}
  M.~Dine, G.~Festuccia, J.~Kehayias, W.~Wu,
  Axions in the Landscape and String Theory,
  JHEP {1101} (2011) 012
  [arXiv:1010.4803 [hep-th]].

\bibitem{Banks:2003es}
  T.~Banks, M.~Dine, E.~Gorbatov,
  Is there a String Theory Landscape?,
  JHEP {0408} (2004) 058
  [hep-th/0309170].
  
\bibitem{Donoghue:2003vs}
  J.~F.~Donoghue,
  Dynamics of M Theory Vacua,
  Phys.\ Rev.\ D {69} (2004) 106012
   [Erratum-ibid.\ D {69} (2004) 129901]
  [hep-th/0310203].

\bibitem{Choi:2009jt}
  K.~-S.~Choi, H.~P.~Nilles, S.~Ramos-Sanchez, P.~K.~S.~Vaudrevange,
  Accions,
  Phys.\ Lett.\ B {675} (2009) 381
  [arXiv:0902.3070 [hep-th]].

\bibitem{Chatzistavrakidis:2012bb}
  A.~Chatzistavrakidis, E.~Erfani, H.~P.~Nilles, I.~Zavala,
  Axiology,
  JCAP {1209} (2012) 006
  [arXiv:1207.1128 [hep-ph]].

\bibitem{Dienes:1996zr}
  K.~R.~Dienes, C.~F.~Kolda, J.~March-Russell,
  Kinetic Mixing and the Supersymmetric Gauge Hierarchy,
  Nucl.\ Phys.\ B {492} (1997) 104
  [hep-ph/9610479].

\bibitem{Goodsell:2011wn}
  M.~Goodsell, S.~Ramos-Sanchez, A.~Ringwald,
  Kinetic Mixing of U(1)s in Heterotic Orbifolds,
  JHEP {1201} (2012) 021
  [arXiv:1110.6901 [hep-th]].

\bibitem{Abel:2008ai}
  S.~A.~Abel, M.~D.~Goodsell, J.~Jaeckel, V.~V.~Khoze, A.~Ringwald,
  Kinetic Mixing of the Photon with Hidden U(1)s in String Phenomenology,
  JHEP {0807} (2008) 124
  [arXiv:0803.1449 [hep-ph]].
  
\bibitem{Goodsell:2009xc}
  M.~Goodsell, J.~Jaeckel, J.~Redondo, A.~Ringwald,
  Naturally Light Hidden Photons in LARGE Volume String Compactifications,
  JHEP {0911} (2009) 027
  [arXiv:0909.0515 [hep-ph]].
  
\bibitem{Cicoli:2011yh}
  M.~Cicoli, M.~Goodsell, J.~Jaeckel, A.~Ringwald,
  Testing String Vacua in the Lab: From a Hidden CMB to Dark Forces in Flux Compactifications,
  JHEP {1107} (2011) 114
  [arXiv:1103.3705 [hep-th]].

\bibitem{Raffelt:1996wa}
  G.~G.~Raffelt,
  Stars as Laboratories for Fundamental Physics: The Astrophysics of Neutrinos, Axions, and Other Weakly Interacting Particles,
  University Press, Chicago, 1996. 

\bibitem{Friedland:2012hj}
  A.~Friedland, M.~Giannotti, M.~Wise,
  ``Constraining the Axion-Photon Coupling with Massive Stars,''
  arXiv:1210.1271 [hep-ph].

\bibitem{Corsico:2001be}
  A.~H.~Corsico, O.~G.~Benvenuto, L.~G.~Althaus, J.~Isern, E.~Garcia-Berro,
  The Potential of the Variable DA White Dwarf G117 - B15A as a Tool for Fundamental Physics,
  New Astron.\  {6} (2001) 197
  [astro-ph/0104103].

\bibitem{Isern:2003xj}
  J.~Isern, E.~Garcia-Berro,
  White Dwarf Stars as Particle Physics Laboratories,
  Nucl.\ Phys.\ Proc.\ Suppl.\  {114} (2003) 107.

\bibitem{Corsico:2012ki}
  A.~H.~Corsico, L.~G.~Althaus, M.~M.~M.~Bertolami, A.~D.~Romero, E.~Garcia-Berro, J.~Isern, S.~O.~Kepler,
  The Rate of Cooling of the Pulsating White Dwarf Star G117$-$B15A: a New Asteroseismological Inference of the Axion Mass,
  arXiv:1205.6180 [astro-ph.SR].

\bibitem{Isern:2008nt}
  J.~Isern, E.~Garcia-Berro, S.~Torres, S.~Catalan,
  Axions and the Cooling of White Dwarf Stars,
  Astrophys.\ J.\  {682} (2008) L109
  [arXiv:0806.2807 [astro-ph]].

\bibitem{Isern:2008fs}
  J.~Isern, S.~Catalan, E.~Garcia-Berro, S.~Torres,
  Axions and the White Dwarf Luminosity Function,
  J.\ Phys.\ Conf.\ Ser.\  {172} (2009)  012005
  [arXiv:0812.3043 [astro-ph]].

\bibitem{Isern:2012ef}
  J.~Isern, L.~Althaus, S.~Catalan, A.~Corsico, E.~Garcia-Berro, M.~Salaris, S.~Torres,
  White Dwarfs as Physics Laboratories: the Case of Axions,
  arXiv:1204.3565 [astro-ph.SR].


\bibitem{Redondo:2008ec}
  J.~Redondo, M.~Postma,
  Massive Hidden Photons as Lukewarm Dark Matter,
  JCAP {0902} (2009) 005
  [arXiv:0811.0326 [hep-ph]].

\bibitem{Redondo:2008aa}
  J.~Redondo,
  Helioscope Bounds on Hidden Sector Photons,
  JCAP {0807} (2008) 008
  [arXiv:0801.1527 [hep-ph]].
  

\bibitem{Brockway:1996yr}
  J.~W.~Brockway, E.~D.~Carlson, G.~G.~Raffelt,
  SN1987A Gamma-Ray Limits on the Conversion of Pseudoscalars,
  Phys.\ Lett.\ B {383} (1996) 439
  [astro-ph/9605197].

\bibitem{Grifols:1996id}
  J.~A.~Grifols, E.~Masso, R.~Toldra,
  Gamma-rays from SN1987A due to Pseudoscalar Conversion,
  Phys.\ Rev.\ Lett.\  {77} (1996) 2372
  [astro-ph/9606028].

\bibitem{Gill:2011yp}
  R.~Gill, J.~S.~Heyl,
  Constraining the Photon-Axion Coupling Constant with Magnetic White Dwarfs,
  Phys.\ Rev.\ D {84} (2011) 085001
  [arXiv:1105.2083 [astro-ph.HE]].

\bibitem{Horns:2012pp}
  D.~Horns, L.~Maccione, A.~Mirizzi, M.~Roncadelli,
  Probing Axion-Like Particles with the Ultraviolet Photon Polarization from Active Galactic Nuclei in Radio Galaxies,
  Phys.\ Rev.\ D {85} (2012) 085021
  [arXiv:1203.2184 [astro-ph.HE]].

\bibitem{Horns:2012fx}
  D.~Horns, M.~Meyer,
  Indications for a Pair-Production Anomaly from the Propagation of VHE Gamma-Rays,
  JCAP {1202} (2012) 033
  [arXiv:1201.4711 [astro-ph.CO]].


\bibitem{DeAngelis:2007dy}
  A.~De Angelis, O.~Mansutti, M.~Roncadelli,
  Evidence for a New Light Spin-Zero Boson from Cosmological Gamma-Ray Propagation?,
  Phys.\ Rev.\ D {76} (2007)  121301.
  [arXiv:0707.4312 [astro-ph]].

\bibitem{Simet:2007sa}
  M.~Simet, D.~Hooper, P.~D.~Serpico,
  The Milky Way as a Kiloparsec-Scale Axionscope,
  Phys.\ Rev.\  D {77} (2008)  063001.
  [arXiv:0712.2825 [astro-ph]].

\bibitem{SanchezConde:2009wu}
  M.~A.~Sanchez-Conde, D.~Paneque, E.~Bloom, F.~Prada, A.~Dominguez,
  Hints of the Existence of Axion-Like-Particles from the Gamma-Ray Spectra of Cosmological Sources,
  Phys.\ Rev.\  D {79} (2009)  123511.
  [arXiv:0905.3270 [astro-ph.CO]].

\bibitem{Horns:2012kw}
  D.~Horns, L.~Maccione, M.~Meyer, A.~Mirizzi, D.~Montanino, M.~Roncadelli,
  Hardening of TeV Gamma Spectrum of AGNs in Galaxy Clusters by Conversions of Photons into Axion-Like Particles,
  arXiv:1207.0776 [astro-ph.HE].

\bibitem{Nakayama:2010vs}
  K.~Nakayama, F.~Takahashi, T.~T.~Yanagida,
  A Theory of Extra Radiation in the Universe,
  Phys.\ Lett.\ B {697} (2011) 275
  [arXiv:1010.5693 [hep-ph]].


\bibitem{Melchiorri:2007sq}
  A.~Melchiorri, A.~Polosa, A.~Strumia,
  New Bounds on Millicharged Particles from Cosmology,
  Phys.\ Lett.\ B {650} (2007) 416
  [hep-ph/0703144].
  
\bibitem{Jaeckel:2008fi}
  J.~Jaeckel, J.~Redondo, A.~Ringwald,
  Signatures of a Hidden Cosmic Microwave Background,
  Phys.\ Rev.\ Lett.\  {101} (2008) 131801
  [arXiv:0804.4157 [astro-ph]].
  
\bibitem{Mirizzi:2009iz}
  A.~Mirizzi, J.~Redondo, G.~Sigl,
  Microwave Background Constraints on Mixing of Photons with Hidden Photons,
  JCAP {0903} (2009) 026
  [arXiv:0901.0014 [hep-ph]].

\bibitem{Cicoli:2012aq}
  M.~Cicoli, J.~P.~Conlon, F.~Quevedo,
  Dark Radiation in LARGE Volume Models,
  arXiv:1208.3562 [hep-ph].
  
\bibitem{Higaki:2012ar}
  T.~Higaki, F.~Takahashi,
  Dark Radiation and Dark Matter in Large Volume Compactifications,
  arXiv:1208.3563 [hep-ph].


\bibitem{Komatsu:2010fb}
  E.~Komatsu {\it et al.}  [WMAP Collaboration],
  Seven-Year Wilkinson Microwave Anisotropy Probe (WMAP) Observations: Cosmological Interpretation,
  Astrophys.\ J.\ Suppl.\  {192} (2011) 18
  [arXiv:1001.4538 [astro-ph.CO]].

\bibitem{Dunkley:2010ge}
  J.~Dunkley
  {\it et al.} [Atacama Cosmology Telescope Collaboration], Cosmological Parameters from the 2008 Power Spectra,
  Astrophys.\ J.\  {739} (2011) 52
  [arXiv:1009.0866 [astro-ph.CO]].


\bibitem{Preskill:1982cy}
  J.~Preskill, M.~B.~Wise, F.~Wilczek,
  Cosmology of the Invisible Axion,
  Phys.\ Lett.\  B {120} (1983) 127.

\bibitem{Abbott:1982af}
  L.~F.~Abbott, P.~Sikivie,
  A Cosmological Bound on the Invisible Axion,
  Phys.\ Lett.\  B {120} (1983) 133.

\bibitem{Dine:1982ah}
  M.~Dine, W.~Fischler,
  The Not So Harmless Axion,
  Phys.\ Lett.\  B {120} (1983) 137.

\bibitem{Sikivie:2006ni}
  P.~Sikivie,
  Axion Cosmology,
  Lect.\ Notes Phys.\  {741} (2008) 19
  [astro-ph/0610440].

\bibitem{Nelson:2011sf}
  A.~E.~Nelson, J.~Scholtz,
  Dark Light, Dark Matter and the Misalignment Mechanism,
  Phys.\ Rev.\ D {84} (2011) 103501
  [arXiv:1105.2812 [hep-ph]].

\bibitem{Tegmark:2005dy}
  M.~Tegmark, A.~Aguirre, M.~Rees, F.~Wilczek,
  Dimensionless Constants, Cosmology and Other Dark Matters,
  Phys.\ Rev.\ D {73} (2006) 023505
  [astro-ph/0511774].

\bibitem{Mack:2009hv} 
  K.~J.~Mack,
  Axions, Inflation and the Anthropic Principle,
  JCAP {1107}  (2011) 021 
  [arXiv:0911.0421 [astro-ph.CO]].

\bibitem{Beltran:2006sq}
  M.~Beltran, J.~Garcia-Bellido, J.~Lesgourgues,
  Isocurvature Bounds on Axions Revisited,
  Phys.\ Rev.\ D {75} (2007) 103507
  [hep-ph/0606107].
  
 
\bibitem{Hertzberg:2008wr}
  M.~P.~Hertzberg, M.~Tegmark, F.~Wilczek,
  Axion Cosmology and the Energy Scale of Inflation,
  Phys.\ Rev.\ D {78} (2008) 083507
  [arXiv:0807.1726 [astro-ph]].
  
  
\bibitem{Hamann:2009yf}
  J.~Hamann, S.~Hannestad, G.~G.~Raffelt, Y.~Y.~Y.~Wong,
  Isocurvature Forecast in the Anthropic Axion Window,
  JCAP {0906} (2009) 022
  [arXiv:0904.0647 [hep-ph]].
  

\bibitem{Wantz:2009it}
  O.~Wantz, E.~P.~S.~Shellard,
  Axion Cosmology Revisited,
  Phys.\ Rev.\ D {82} (2010) 123508
  [arXiv:0910.1066 [astro-ph.CO]].

\bibitem{Hiramatsu:2010yu}
  T.~Hiramatsu, M.~Kawasaki, T.~Sekiguchi, M.~Yamaguchi, J.~'i.~Yokoyama,
  Improved Estimation of Radiated Axions from Cosmological Axionic Strings,
  Phys.\ Rev.\ D {83} (2011) 123531
  [arXiv:1012.5502 [hep-ph]].
  
\bibitem{Hiramatsu:2012gg}
  T.~Hiramatsu, M.~Kawasaki, K.~'i.~Saikawa, T.~Sekiguchi,
  Production of Dark Matter Axions from Collapse of String-Wall Systems,
  Phys.\ Rev.\ D {85} (2012) 105020
  [arXiv:1202.5851 [hep-ph]].

\bibitem{Kolb:1993zz}
  E.~W.~Kolb, I.~I.~Tkachev,
  Axion Miniclusters and Bose Stars,
  Phys.\ Rev.\ Lett.\  {71} (1993) 3051
  [hep-ph/9303313].
  
\bibitem{Zurek:2006sy}
  K.~M.~Zurek, C.~J.~Hogan, T.~R.~Quinn,
  Astrophysical Effects of Scalar Dark Matter Miniclusters,
  Phys.\ Rev.\ D {75} (2007) 043511
  [astro-ph/0607341].

\bibitem{Choi:2008zq}
  K.~-Y.~Choi, J.~E.~Kim, H.~M.~Lee, O.~Seto,
  Neutralino Dark Matter from Heavy Axino Decay,
  Phys.\ Rev.\ D {77} (2008) 123501
  [arXiv:0801.0491 [hep-ph]].

\bibitem{Baer:2011hx}
  H.~Baer, A.~Lessa, S.~Rajagopalan, W.~Sreethawong,
  Mixed Axion/Neutralino Cold Dark Matter in Supersymmetric Models,
  JCAP {1106} (2011) 031
  [arXiv:1103.5413 [hep-ph]].

\bibitem{Baer:2011uz}
  H.~Baer, A.~Lessa, W.~Sreethawong,
  Coupled Boltzmann Calculation of Mixed Axion/Neutralino Cold Dark Matter Production in the Early Universe,
  JCAP {1201} (2012) 036
  [arXiv:1110.2491 [hep-ph]].

\bibitem{Sikivie:2009qn}
  P.~Sikivie, Q.~Yang,
  Bose-Einstein Condensation of Dark Matter Axions,
  Phys.\ Rev.\ Lett.\  {103} (2009) 111301
  [arXiv:0901.1106 [hep-ph]].

\bibitem{Erken:2011dz}
  O.~Erken, P.~Sikivie, H.~Tam, Q.~Yang,
  Cosmic Axion Thermalization,
  Phys.\ Rev.\ D {85} (2012) 063520
  [arXiv:1111.1157 [astro-ph.CO]].

\bibitem{Sikivie:2010yn}
  P.~Sikivie,
  The Dark Matter is Mostly an Axion BEC,
  PoS IDM {2010} (2011) 068
  [arXiv:1012.1553 [astro-ph.CO]].

\bibitem{Erken:2011xj}
  O.~Erken, P.~Sikivie, H.~Tam, Q.~Yang,
  Axion BEC Dark Matter,
  arXiv:1111.3976 [astro-ph.CO].

\bibitem{Kain:2011pd}
  B.~Kain, H.~Y.~Ling,
  Cosmological Inhomogeneities with Bose-Einstein Condensate Dark Matter,
  Phys.\ Rev.\ D {85} (2012) 023527
  [arXiv:1112.4169 [hep-ph]].

\bibitem{Sikivie:2012gi}
  P.~Sikivie,
  An Argument that the Dark Matter is Axions,
  arXiv:1210.0040 [astro-ph.CO].

\bibitem{Dobrich:2010hi}
  B.~Dobrich, H.~Gies,
  Axion-like-particle search with high-intensity lasers,
  JHEP {1010} (2010) 022
  [arXiv:1006.5579 [hep-ph]].

\bibitem{Dobrich:2010ie}
  B.~Dobrich, H.~Gies,
  High-Intensity Probes of Axion-Like Particles,
  arXiv:1010.6161 [hep-ph].

\bibitem{Sikivie:1983ip}
  P.~Sikivie,
  Experimental Tests of the Invisible Axion,
  Phys.\ Rev.\ Lett.\  {51} (1983) 1415
  [Erratum-ibid.\  {52} (1984) 695].
  
\bibitem{DePanfilis:1987dk}
  S.~De Panfilis 
  {\it et al.},
  Limits on the Abundance and Coupling of Cosmic Axions at 4.5 Micro-eV $<$ m(a) $<$ 5.0 Micro-eV,
  Phys.\ Rev.\ Lett.\  {59} (1987) 839.

\bibitem{Wuensch:1989sa}
  W.~Wuensch 
  {\it et al.},
  Results of a Laboratory Search for Cosmic Axions and Other Weakly Coupled Light Particles,
  Phys.\ Rev.\ D {40} (1989) 3153.

\bibitem{Hagmann:1990tj}
  C.~Hagmann, P.~Sikivie, N.~S.~Sullivan, D.~B.~Tanner,
  Results from a Search for Cosmic Axions,
  Phys.\ Rev.\ D {42} (1990) 1297.

\bibitem{Asztalos:2001tf}
  S.~J.~Asztalos 
  {\it et al.},
  Large Scale Microwave Cavity Search for Dark Matter Axions,
  Phys.\ Rev.\ D {64} (2001) 092003.

\bibitem{Asztalos:2009yp}
  S.~J.~Asztalos {\it et al.}  [The ADMX Collaboration],
  A SQUID-based Microwave Cavity Search for Dark-Matter Axions,
  Phys.\ Rev.\ Lett.\  {104} (2010) 041301
  [arXiv:0910.5914 [astro-ph.CO]].


\bibitem{Asztalos:2011bm}
  S.~J.~Asztalos 
  {\it et al.},
  Design and Performance of the ADMX SQUID-Based Microwave Receiver,
  Nucl.\ Instrum.\ Meth.\ A {656} (2011) 39
  [arXiv:1105.4203 [physics.ins-det]].

\bibitem{Heilman:2010zz}
  J.~Heilman {\it et al.}  [ADMX Collaboration],
  ADMX phase II: Relocation and Millikelvin cooling,
  AIP Conf.\ Proc.\  {1274} (2010) 115.

\bibitem{Baker:2011na}
  O.~K.~Baker 
  {\it et al.},
  Prospects for Searching Axion-like Particle Dark Matter with Dipole, Toroidal and Wiggler Magnets,
  Phys.\ Rev.\ D {85} (2012) 035018
  [arXiv:1110.2180 [physics.ins-det]].

\bibitem{Melissinos:2008vn}
  A.~C.~Melissinos,
  Search for Cosmic Axions using an Optical Interferometer,
  Phys.\ Rev.\ Lett.\  {102} (2009) 202001
  [arXiv:0807.1092 [hep-ph]].

\bibitem{Graham:2011qk}
  P.~W.~Graham, S.~Rajendran,
  Axion Dark Matter Detection with Cold Molecules,
  Phys.\ Rev.\ D {84} (2011) 055013
  [arXiv:1101.2691 [hep-ph]].

\bibitem{Horns:inprep}
D. Horns, J. Jaeckel, A. Lindner, A. Lobanov, J. Redondo, A. Ringwald, 
Feasibility of Broadband Radio Searches of WISPy Cold Dark Matter, 
in preparation.


\bibitem{Andriamonje:2007ew}
  S.~Andriamonje {\it et al.}  [CAST Collaboration],
  An Improved Limit on the Axion-Photon Coupling from the CAST Experiment,
  JCAP {0704} (2007) 010
  [hep-ex/0702006].

\bibitem{Arik:2008mq}
  E.~Arik {\it et al.}  [CAST Collaboration],
  Probing eV-Scale Axions with CAST,
  JCAP {0902} (2009) 008
  [arXiv:0810.4482 [hep-ex]].

\bibitem{Arik:2011rx}
  S.~Aune {\it et al.}  [CAST Collaboration],
  CAST Search for Sub-eV Mass Solar Axions with 3He Buffer Gas,
  Phys.\ Rev.\ Lett.\  {107} (2011) 261302
  [arXiv:1106.3919 [hep-ex]].

\bibitem{Irastorza:2011gs}
  I.~G.~Irastorza 
  {\it et al.},
  Towards a New Generation Axion Helioscope,
  JCAP {1106} (2011) 013
  [arXiv:1103.5334 [hep-ex]].

\bibitem{Schwarz:2011gu}
  M.~Schwarz, A.~Lindner, J.~Redondo, A.~Ringwald, G.~Wiedemann,
  Solar Hidden Photon Search,
  arXiv:1111.5797 [astro-ph.IM].

\bibitem{Anselm:1986gz}
  A.~A.~Anselm,
  Arion <---> Photon Oscillations in a Steady Magnetic Field,
  Yad.\ Fiz.\  {42} (1985) 1480.
  
\bibitem{VanBibber:1987rq}
  K.~Van Bibber, N.~R.~Dagdeviren, S.~E.~Koonin, A.~Kerman, H.~N.~Nelson,
  Proposed Experiment to Produce and Detect Light Pseudoscalars,
  Phys.\ Rev.\ Lett.\  {59} (1987) 759.
  
\bibitem{Redondo:2010dp}
  J.~Redondo, A.~Ringwald,
  Light Shining Through Walls,
  Contemp.\ Phys.\  {52} (2011) 211
  [arXiv:1011.3741 [hep-ph]].

\bibitem{Okun:1982xi}
  L.~B.~Okun,
  Limits Of Electrodynamics: Paraphotons?,
  Sov.\ Phys.\ JETP {56} (1982) 502
   [Zh.\ Eksp.\ Teor.\ Fiz.\  { 83} (1982) 892].

\bibitem{Ahlers:2007rd}
  M.~Ahlers, H.~Gies, J.~Jaeckel, J.~Redondo, A.~Ringwald,
  Light from the Hidden Sector,
  Phys.\ Rev.\ D {76} (2007) 115005
  [arXiv:0706.2836 [hep-ph]].

\bibitem{Ahlers:2007qf}
  M.~Ahlers, H.~Gies, J.~Jaeckel, J.~Redondo, A.~Ringwald,
  Laser Experiments Explore the Hidden Sector,
  Phys.\ Rev.\ D {77} (2008) 095001
  [arXiv:0711.4991 [hep-ph]].

\bibitem{Dobrich:2012sw}
  B.~Dobrich, H.~Gies, N.~Neitz, F.~Karbstein, 
  Magnetically Amplified Tunneling of the 3rd Kind as a Probe of Minicharged Particles,
  Phys.\ Rev.\ Lett.\  {109} (2012) 131802
  [arXiv:1203.2533 [hep-ph]].

\bibitem{Dobrich:2012jd}
  B.~Dobrich, H.~Gies, N.~Neitz, F.~Karbstein,
  Magnetically Amplified Light-Shining-Through-Walls via Virtual Minicharged Particles,
  arXiv:1203.4986 [hep-ph].

\bibitem{Jaeckel:2006xm}
  J.~Jaeckel, E.~Masso, J.~Redondo, A.~Ringwald, F.~Takahashi,
  The Need for Purely Laboratory-Based Axion-Like Particle Searches,
  Phys.\ Rev.\ D {75} (2007) 013004
  [hep-ph/0610203].

\bibitem{Ehret:2009sq}
  K.~Ehret {\it et al.}  [ALPS Collaboration],
  Resonant Laser Power Build-Up in ALPS: A 'Light-Shining-Through-Walls' Experiment,
  Nucl.\ Instrum.\ Meth.\ A {612} (2009) 83
  [arXiv:0905.4159 [physics.ins-det]].
 
\bibitem{Ehret:2010mh}
  K.~Ehret 
  {\it et al.},
  New ALPS Results on Hidden-Sector Lightweights,
  Phys.\ Lett.\ B {689} (2010) 149
  [arXiv:1004.1313 [hep-ex]].

\bibitem{alpsiitdr}
  [ALPS Collaboration],
  Any Light Particle Search II, Technical Design Report,
  2012.
  
\bibitem{Ringwald:2003nsa}
  A.~Ringwald,
  Production and Detection of Very Light Bosons in the HERA Tunnel,
  Phys.\ Lett.\ B {569} (2003) 51
  [hep-ph/0306106].

\bibitem{Hoogeveen:1990vq}
  F.~Hoogeveen, T.~Ziegenhagen,
  Production and Detection of Light Bosons Using Optical Resonators,
  Nucl.\ Phys.\ B {358} (1991) 3.

\bibitem{Mueller:2009wt}
  G.~Mueller, P.~Sikivie, D.~B.~Tanner, K.~van Bibber,
  Detailed Design of a Resonantly-Enhanced Axion-Photon Regeneration Experiment,
  Phys.\ Rev.\ D {80} (2009) 072004
  [arXiv:0907.5387 [hep-ph]].

\bibitem{Hoogeveen:1992nq}
  F.~Hoogeveen,
  Terrestrial Axion Production and Detection Using RF Cavities,
  Phys.\ Lett.\ B {288} (1992) 195.

\bibitem{Jaeckel:2007ch}
  J.~Jaeckel, A.~Ringwald,
  A Cavity Experiment to Search for Hidden Sector Photons,
  Phys.\ Lett.\ B {659} (2008) 509
  [arXiv:0707.2063 [hep-ph]].
  
\bibitem{Caspers:2009cj}
  F.~Caspers, J.~Jaeckel, A.~Ringwald,
  Feasibility, Engineering Aspects and Physics Reach of Microwave Cavity Experiments Searching for Hidden Photons and Axions,
  JINST {4} (2009) P11013
  [arXiv:0908.0759 [hep-ex]].

\bibitem{Rabadan:2005dm}
  R.~Rabadan, A.~Ringwald, K.~Sigurdson,
  Photon Regeneration from Pseudoscalars at X-ray Laser Facilities,
  Phys.\ Rev.\ Lett.\  {96} (2006) 110407
  [hep-ph/0511103].

\bibitem{Dias:2009ph}
  A.~G.~Dias, G.~Lugones,
  Probing Light Pseudoscalar Particles Using Synchrotron Light,
  Phys.\ Lett.\ B {673} (2009) 101
  [arXiv:0902.0749 [hep-ph]].

\bibitem{Wagner:2010mi}
  A.~Wagner 
  {\it et al.},
  A Search for Hidden Sector Photons with ADMX,
  Phys.\ Rev.\ Lett.\  {105} (2010) 171801
  [arXiv:1007.3766 [hep-ex]].

\bibitem{Betz:2012tp}
  M.~Betz, F.~Caspers,
  A Microwave Paraphoton and Axion Detection Experiment with 300 dB Electromagnetic Shielding at 3 GHz,
  Conf.\ Proc.\ C {1205201} (2012) 3320
  [arXiv:1207.3275 [physics.ins-det]].



\bibitem{Battesti:2010dm}
  R.~Battesti 
  {\it et al.},
  A Photon Regeneration Experiment for Axionlike Particle Search using X-rays,
  Phys.\ Rev.\ Lett.\  {105} (2010) 250405
  [arXiv:1008.2672 [hep-ex]].

\end{thebibliography}



\end{document}